\documentclass[twocolumn,english,aps,prx,floatfix, amsfonts]{revtex4-1}
\usepackage{epsfig, graphicx,psfrag,amsmath,amssymb,float}
\usepackage[T1]{fontenc}
\usepackage[latin9]{inputenc}
\usepackage{amsmath}
\usepackage{amssymb}
\usepackage{amscd}
\usepackage{bm}
\usepackage{psfrag}
\usepackage{babel}
\usepackage{mathrsfs}

\usepackage{color}
\usepackage[bookmarks=true,colorlinks,linkcolor=blue,urlcolor=blue,citecolor=blue]{hyperref}

\newcommand{\be}{\begin{equation}}
\newcommand{\ee}{\end{equation}}
\newcommand{\bea}{\begin{eqnarray}}
\newcommand{\eea}{\end{eqnarray}}

\newcommand{\mc}{\mathcal}
\newcommand{\mb}{\mathbf}

\newcommand{\added}[1]{\textcolor{magenta}{#1}}
\begin{document}

\title{Superconductivity from piezoelectric interactions in Weyl semimetals}

\author{Rodrigo G.~Pereira,${}^{1,2}$ Francesco Buccheri,${}^2$ Alessandro De Martino,${}^3$ and Reinhold Egger${}^2$\ }
\affiliation{${}^1$~International Institute of Physics and Departamento de F{\'i}sica Te{\'o}rica e Experimental, Universidade Federal do Rio Grande do Norte, Campus Universitario, Lagoa Nova, Natal-RN 59078-970, Brazil\\
${}^2$~Institut f\"ur Theoretische Physik, Heinrich-Heine-Universit\"at, D-40225  D\"usseldorf, Germany, \\
${}^3$~Department of Mathematics, City, University of London, EC1V 0HB London, UK}

\date{\today}
\begin{abstract}
We present an analytical low-energy theory of piezoelectric electron-phonon interactions in undoped Weyl semimetals, taking into account also  Coulomb interactions. We show that piezoelectric interactions generate a long-range attractive potential between Weyl fermions. This potential comes with a characteristic angular anisotropy.  From the one-loop renormalization group approach and a mean-field analysis, we predict that superconducting phases with either conventional $s$-wave singlet pairing or nodal-line triplet pairing could be realized for sufficiently strong piezoelectric coupling.  For small
couplings, we show that the quasi-particle decay rate exhibits a linear temperature dependence where the prefactor vanishes only in a  logarithmic manner as the quasi-particle energy approaches the Weyl point.  For practical estimates, we consider the Weyl semimetal TaAs.      
\end{abstract}

\maketitle

\section{Introduction}  \label{sec1}

The physics of three-dimensional (3D) Weyl semimetals (WSMs) is presently attracting a lot of interest. For several different candidate materials, experiments have recently revealed WSM signatures in various observables  \cite{Jia2016,Yan2017,Hasan2017}.  Within band theory, WSMs have an even number of touching points (the so-called Weyl nodes) in the Brillouin zone. Near those special points, low-energy quasi-particles have a linear spectrum and represent Weyl fermions \cite{Volovik,Burkov2016,Hosur2013,Burkov2015,Burkov2018,Armitage2018}. 
The Weyl character of low-energy fermions implies the existence of a chiral anomaly which in turn produces characteristic signatures in  experimentally accessible observables such as the magnetoconductivity \cite{Burkov2018}.  The remarkable transport features of WSMs may also lead to useful practical applications  \cite{Ali,Parameswaran}.

We here study the theory of electron-phonon (e-ph) interactions in WSMs.  Apart from the case of optical phonons \cite{Song2016,Rinkel2017,Liu2017,Gordon2018,Rinkel,new19}, the exploration of  e-ph coupling effects in WSMs has not received much attention by theorists so far. However, it has been pointed out that in the static (frozen phonon) limit, 
strain engineering can be used to induce pseudo-scalar and pseudo-vector potentials that couple to Weyl fermions \cite{Liu2013,Cortijo2015,Shapourian2015,Pikulin2016,Grushin2016,Moeller2017,Ferreiros2019}.  
We here focus on low-energy long-wavelength acoustic phonons with linear dispersion, schematically written as $\Omega(\mb q)=c_{ph}|\mb q|$ with sound velocity $c_{ ph}$. The linear dispersion of phonons as well as Weyl fermions suggests the existence of a scale-invariant effective action that may allow for nontrivial fixed points under the renormalization group (RG). We shall assume below that all relevant phonon momenta are well below the momentum separation $b$ between a time-reversed pair of Weyl nodes, $|\mb q|\ll b$, such that phonons cannot scatter electrons between Weyl points at low temperatures. However, at elevated temperatures, $T\agt c_{ph} b/k_B$,  this assumption breaks down and additional processes not considered in this work could take place.

For insulators or semiconductors,  the most important couplings between electrons and acoustic phonons generally originate from either the deformation potential or the piezoelectric interaction \cite{MahanBook,Yu,Giustino}. While the former is a short-range interaction, the latter represents an anisotropic  long-range interaction that only exists for inversion-symmetry-breaking crystals. 
The so-called direct piezoelectric effect   refers to the appearance of an electric polarization when a material is subjected to static stress. 
On the other hand, in a metal, free charge carriers will screen the electric fields produced by local dipole moments, thereby preventing any macroscopic polarization. Nonetheless, it is still possible to speak of piezoelectricity in metals by measuring the bulk electric current in response to a time-dependent strain \cite{Varjas,Vanderbilt}.  Electric currents in response to strain have been discussed in the context of WSMs in Ref.~\cite{Cortijo}.   
Below we will employ piezoelectric coupling expressions derived within the phenomenological theory of electronic insulators \cite{Mahan}. The main assumptions behind this approach are that the electric field produced by phonons is approximately longitudinal, and that there are no free charge carriers responsible for screening. In that case, $\nabla\cdot \mb D=0$ can be assumed for the electric displacement field $\mb D$. 
A microscopic derivation of the piezoelectric coupling \cite{Vogl} gives further support to this phenomenological theory. 
The microscopic approach directly applies to insulators, where one can neglect the frequency dependence of the permittivity  
at frequencies well below the energy gap. 

In \emph{undoped} WSMs, the Fermi level is aligned with a Weyl point. Albeit the spectrum is gapless, screening is absent since  the density of states vanishes at the Weyl point even when
 weak disorder is taken into account \cite{Altland2018}.   In fact, electron-electron (e-e) interactions are marginally irrelevant in 3D WSMs, such that the dielectric function picks up only logarithmic corrections at low energy scales \cite{Abrikosov,Isobe1,Isobe,Yang,Throckmorton}.  However,   when  computing finite-temperature observables, it may be necessary to include the dynamic screening effects represented by  these  logarithmic corrections, as we will discuss in Sec.~\ref{sec4c} in more detail.

We thus conclude that the piezoelectric coupling in undoped WSMs 
can be obtained along the lines of Refs.~\cite{Mahan,Giustino,Vogl}, see Eq.~\eqref{piezoqdep} below.
If piezoelectric couplings are finite, we find that they dominate over all other types of e-ph couplings which 
represent RG-irrelevant short-range interactions. 
Since many  WSMs  discovered so far  belong to polar crystal symmetry classes, e.g., the  ditetragonal-pyramidal $4mm$ class for TaAs, piezoelectric couplings are expected to play an important role for a wide class of WSM materials.  Our general results will below be illustrated for the concrete case of TaAs, which also represents one of the experimentally most intensely studied WSMs  
\cite{Xu2015,Lv2015,Yang2015,Lv2015b,Huang2015,Zhang2016,Zhou2016,Arnold2016,Xu2017,Zhang2017}.
For related \emph{ab initio} results, see Refs.~\cite{Huang2015b,Buckeridge}.

In this paper, we present an analytical  theory capturing the generic low-energy physics 
of undoped 3D WSMs taking into account the piezoelectric e-ph interaction.
We also include e-e interactions even though they represent marginally irrelevant perturbations   in  WSMs. Nonetheless, their interplay with the piezoelectric coupling  may lead to an instability  in the RG flow \cite{Cardy} which  
drives the WSM into a Weyl superconductor  \cite{Armitage2018,Meng,Cho,Wei,Hosur,Bednik,Li,Gorbar2019} phase. For a similar but different study of e-e and e-ph interactions in the context of two-dimensional (2D) Dirac fermions in graphene layers, see Ref.~\cite{Basko}.  The main limitations of our theory come from the neglect of disorder and from the often rather complex band structure of real WSM materials.  Moreover, we confine ourselves to \emph{bulk} properties only, leaving studies of surface state properties to future research.

The structure of the remainder of this paper is as follows. In Sec.~\ref{sec2}, we explain the model used in our study, derive the piezoelectric coupling Hamiltonian, and introduce a local field theory capturing both e-e and e-ph interactions.   We use this field theory to derive the effective interaction potential between two Weyl fermions and show that the phonon-mediated attractive contribution has a characteristic angular anisotropy.  In Sec.~\ref{sec4a}, we   provide parameter estimates for the example of TaAs.  
In Sec.~\ref{sec3}, we then derive and discuss the RG equations found from a one-loop analysis.
We continue in Sec.~\ref{sec4} by investigating the stability of different superconducting phases by an analytical mean-field analysis. In addition, \added{in Sec.~\ref{sec4c},} we   address the temperature and momentum dependence of the quasi-particle decay rate for small piezoelectric couplings where no interaction-induced instabilities are expected. Finally, we offer our conclusions in Sec.~\ref{sec5}.  Technical details can be found in the Appendix. We put $\hbar=k_B=1$ throughout. 

\section{Piezoelectric  interactions in Weyl semimetals} \label{sec2}

In this section, we describe the model used in this work and derive the piezoelectric coupling between electrons and acoustic phonons in undoped WSMs.  We first briefly summarize the electronic Weyl Hamiltonian in Sec.~\ref{sec2a}, and then discuss a general acoustic phonon model in Sec.~\ref{sec2b}. We proceed in Sec.~\ref{sec2c} with a derivation of the piezoelectric coupling Hamiltonian. Next, in Sec.~\ref{sec2d}, we introduce a local field theory approach in order to capture both Coulomb interactions
and piezoelectric interactions on equal footing.  We also derive the attractive phonon-mediated potential and show that it exhibits a pronounced angular anisotropy.

\subsection{Weyl Hamiltonian}\label{sec2a}

In the absence of e-e and e-ph interactions, fermionic quasi-particles near a given Weyl node are described by the Weyl Hamiltonian \cite{Volovik,Burkov2016,Hosur2013,Burkov2015,Burkov2018},
\be\label{Hplus}
 H_0 =\sum_{\mb p}\psi^\dagger(\mb p)\left [v_\perp\mb p_\perp\cdot\boldsymbol\sigma_\perp+v_3 p_3 \sigma_3\right]\psi(\mb p),
 \ee
 where the momentum $\mb p=(\mb p_\perp,p_3)$ is measured with respect to the Weyl node, $\psi=(\psi_\uparrow,\psi_\downarrow)^t$ is a spinor field operator, and the Pauli matrices  $\boldsymbol \sigma_\perp=(\sigma_1,\sigma_2)$ and $\sigma_3$ (with identity $\sigma_0$) act in spin space.
 In Eq.~\eqref{Hplus} we consider anisotropic Fermi velocities, $v_3\ne v_\perp$.  In fact, such anisotropies can be generated  by the piezoelectric interaction in crystals with tetragonal symmetry, see Sec.~\ref{sec3a4} below. However, for simplicity, we will often specialize to the isotropic case with
 \be\label{isotropic}
 v_\perp= v_3 = v.
 \end{equation}
 Throughout we assume that the chemical potential is located exactly at the Weyl node.  
 
 WSMs have an even number $2N$ of Weyl nodes in the Brillouin zone.  In particular,
 time-reversal invariant WSMs with at least four Weyl nodes generically appear as intermediate phases between the trivial and the topological insulator phases of non-centrosymmetric semiconductors, where --- depending on the space group of the crystal --- all $2N$ Weyl nodes could be located at the Fermi level \cite{Murakami,Belopolski2017}. For a  continuum model that produces four Weyl nodes by breaking the reflection symmetry of a Dirac semimetal, see Ref.~\cite{Hosur2013}.   
 
Below we employ the fermionic Matsubara Green's function (GF)  \cite{MahanBook,Altland} for Weyl fermions near a given node,
\bea
 G_{\sigma\sigma'}(x-x')=-\langle T_\tau \psi^{\phantom\dagger}_\sigma(x)\psi^{\dagger}_{\sigma'}(x')\rangle,
 \eea
 where $T_\tau$ denotes imaginary time ($\tau$) ordering, the spin index is $\sigma=\uparrow,\downarrow$, and we use the four-vector notation $x=(\tau,\mb r).$
 Taking the Fourier transform,
 with four-momentum $p=(i\omega,\mb p)$,
 the GF has the spin matrix form 
 \be\label{GFT}
\mathbb G(x)=\frac{1}{\beta V}\sum_{p} e^{-i\omega\tau+i\mb p\cdot \mb r}\, \mathbb G(p),
 \ee
  where  $\omega$ denotes fermionic Matsubara frequencies, the volume is $V$, and $\beta=1/T$. Equation~\eqref{Hplus} yields the GF matrix
   \be\label{GF1}
 \mathbb G(p)=\frac{i\omega\sigma_0+v_\perp\mb p_\perp\cdot \boldsymbol\sigma_\perp+v_3p_3\sigma_3}{(i\omega)^2-E^2(\mb p)},
 \ee
 which has poles at $i\omega=\pm E(\mb p)$ with 
 \begin{equation}\label{energy1}
    E(\mb p)=\sqrt{v_\perp^2\mb p_\perp^2+v_3^2p_3^2}.
 \end{equation}
 Such a gapless dispersion relation is characteristic for 3D Weyl fermions.  For the isotropic case (\ref{isotropic}), this yields the 
 familiar massless Weyl fermion dispersion 
 with $E(\mb p)=v|\mb p|$.
Unless noted otherwise, we consider the thermodynamic limit with $T=0$, where all discrete sums such as those appearing in Eq.~\eqref{GFT} are replaced by integrals.
 This step also implies that we investigate only bulk physics.
 
It will sometimes be  advantageous to work in the band basis  where $\mathbb G(p)$ is diagonal. Labeling these bands 
by $\mu=\pm$ and using Eq.~\eqref{energy1}, we find
\be \label{bandGF}
G_{\mu\mu'}(p)=\frac{\delta_{\mu\mu'}}{i\omega-\mu E(\mb p)}\equiv \delta_{\mu\mu'}G_\mu(p).
\ee
The mode expansion for the fermion field then reads
\be \label{bandmode}
\psi_\sigma(\mb r)=\frac1{\sqrt{V}}\sum_{\mb p}\mc U_{\sigma\mu}(\mb p)\psi_{\mb p,\mu}e^{i\mb p\cdot \mb r},
\ee
where $\mc U(\mb p)$ is the unitary matrix that diagonalizes the single-particle Hamiltonian in Eq.~(\ref{Hplus}). Note that $\mc U=\mc U(\hat {\mb p})$ is   a function of the angles
defined by the unit vector $\hat{\mb p}=\mb p/|\mb p|$ in momentum space. The Fourier transform of the electron density operator, $\rho_e(\mb r)=\psi^\dagger\psi$, is then given by
\be\label{charge1}
\rho_e(\mb q)=\sum_{\mb p,\mu,\mu'}\left[\mc U^\dagger(\mb p)\mc U(\mb p+\mb q)\right]_{\mu\mu'}\psi^\dagger_{\mb p\mu}\psi^{\phantom\dagger}_{\mb p+\mb q,\mu'}.
\ee
Allowing  for contributions from all $2N$ Weyl nodes in the Brillouin zone (indexed by $h$), we have $\rho_e(\mb r)=\sum_h\psi^\dagger_h\psi^{\phantom\dagger}_h$.

\subsection{Phonons}\label{sec2b}

We here focus on  acoustic phonons at  long wave lengths.  The physics is then described by the lattice displacement field
$\mb u(\mb r)$. With the linearized strain tensor,
\be \label{strain}
u_{jk}=\frac12(\partial_j u_k+\partial_k u_j),
\ee
and the fourth-order stiffness tensor $C_{ijkl}$, 
 the Euclidean action is given by \cite{MahanBook,Landau}
\be\label{phononaction}
S_{\rm ph}= \int d^4 x \left( \frac{\rho_0}{2} (\partial_\tau {\bf u})^2+ \frac12\sum_{ijkl} C_{ijkl}  u_{ij} u_{kl}\right),
\ee
where $\rho_0$ is the mass  density and $d^4 x=d\tau d^3{\bf r}$. 
Our main interest in this work is in describing possible electronic instabilities of WSMs due to piezoelectric interactions, and we will therefore not study a specific phonon model.
We assume instead that all three ($J=1,2,3)$ acoustic phonon modes have a 
linear dispersion, 
\begin{equation}\label{phonondisp}
    \Omega_J(\mb q)= c_J(\hat{\mb q}) \, |\mb q|,
\end{equation}
where the respective sound velocity, $c_J(\hat{\mb q})$,  could depend on the angular direction  $\hat{\mb q}={\bf q}/|{\bf q}|$.  
Using bosonic annihilation  operators, $a^{}_J(\mb q)$, the standard 
 mode expansion of the lattice displacement field is  given by \cite{MahanBook}
\be
\mb u(\mathbf r)= \sum_{J=1}^3\sum_{\mathbf q} \frac {\boldsymbol\epsilon^J(\mathbf q)e^{i\mathbf q\cdot \mathbf r} }{\sqrt{2\rho_0 V\Omega_J(\mathbf q)}} a^{\phantom\dagger}_J(\mb q) + {\rm h.c.},
\ee
where the $\boldsymbol \epsilon^J(\mb q)$ are polarization unit vectors.

Next we define the phonon propagator   \cite{Altland},
\be
D_{jk}(x-x')=-\langle T_\tau u_j(x)u_{k}(x')\rangle. \label{Djjphonon}
\ee
Taking the Fourier transform and using $q=(i\omega,\mb q)$ with bosonic Matsubara frequencies $\omega$, we obtain from Eqs.~\eqref{phonondisp} and \eqref{Djjphonon} the result
\be
D_{jk}(q)=\frac1{\rho_0}\sum_{J}\frac{\epsilon^J_j(\mb q)\epsilon^J_{k}(-\mb q)}{(i\omega)^2-\Omega^2_J(\mb q)} = D_{kj}(-q).
\label{phonD}
\ee
For an isotropic continuum, we may identify $J=1$ with the longitudinal mode and $J=2,3$ with the transverse modes, where $c_1=c_l$ and $c_{2,3}=c_t$ denote the longitudinal  and transverse sound velocities, respectively.
We will often make the simplifying assumption
\begin{equation}\label{isophon}
    c_t = c_l = c_{ph},\quad c_{ph}\ll v,
\end{equation}
on top of the isotropic Fermi velocity condition \eqref{isotropic}. These assumptions do not affect scaling properties in an essential way.
Moreover, relaxing those approximations does not pose conceptual problems and could allow one to take into account \emph{ab initio} results, see, e.g., Refs.~\cite{Buckeridge,Chang2016}.

\subsection{Piezoelectric interaction}\label{sec2c}

A microscopic derivation of the e-ph interaction in insulators encounters short-range as well as long-range interactions \cite{Vogl,Giustino}. The  long-range contributions can be organized in terms of a multipole expansion of the electron-ion interaction potential. The first term in this expansion is a dipolar contribution which must vanish due to the acoustic sum rule. The next terms are quadrupolar contributions which  account for piezoelectric couplings and vanish for centrosymmetric materials, but not when inversion symmetry is broken. A phenomenological derivation \cite{Mahan,Yu} starts from the constitutive relation for the electric displacement,
\be
D_i=\sum_{jk}e_{ijk}u_{jk}+
\sum_j \varepsilon_{ij}E_j,\label{constitutive}
\ee
where $\mb E$ is the external electric field, $e_{ijk}$ the piezoelectric tensor, and $\varepsilon_{ij}$ the permittivity tensor \cite{MahanBook}. 
A non-vanishing piezoelectric tensor arises if strain can induce $\mb D\ne 0$ even for $\mb E=0$. The relation 
$e_{ijk}=(\partial D_i/\partial u_{jk})_{E}$ and
 the symmetry of the strain tensor, $u_{jk}=u_{kj}$, imply that the piezoelectric tensor is symmetric in the last two indices, $e_{ijk}=e_{ikj}$.
 
 In the absence of free charges, from Eq.~\eqref{constitutive} we  have  
\be
\nabla\cdot \mb D=0=\sum_{ijk}e_{ijk}\partial_iu_{jk}+\sum_{ij}\varepsilon_{ij}\partial_iE_j.
\ee
 Taking the Fourier transform gives
 \be
\sum_{ij}\varepsilon_{ij}q_iE_j(\mb q)=-i\sum_{ijk}e_{ijk}q_iq_ju_{k}(\mb q).
 \ee
Since the electric field is effectively longitudinal \cite{Mahan}, we can write $\mb E(\mb q)\simeq -i\mb q\Phi(\mb q)$ with the scalar potential 
\be\label{scalarpot}
\Phi(\mb q)=\frac{1}{\varepsilon \mb q^2} \sum_{ijk}e_{ijk}q_iq_ju_{k}(\mb q).
\ee
For notational simplicity, we assume an isotropic permittivity tensor, $\varepsilon_{ij}=\varepsilon\delta_{ij}$.

The scalar potential \eqref{scalarpot} now couples to the electronic charge density, cf.~Eq.~\eqref{charge1}, resulting 
in the piezoelectric interaction Hamiltonian
\be\label{piezoqdep}
H_{\textrm{pz}}=\frac{e}{\varepsilon V}\sum_{ijk}\sum_{\mb q\ne 0}e_{ijk}\frac{ q_iq_j}{ \mb q^2} u_k(\mb q)\rho_e(-\mb q).
\ee
We emphasize that the coupling strength in Eq.~\eqref{piezoqdep}  depends on the direction of the unit vector $\hat{\mb q}$,
where the $\mb q=0$ mode 
is omitted  to ensure overall electric neutrality.
From dimensional analysis, $H_{\rm pz}$ is marginal under RG transformations, and second-order perturbation theory implies a linear-in-$T$ dependence of the quasi-particle decay rate, see Sec.~\ref{sec4c} for details.  
At low $T$, the piezoelectric interaction will therefore dominate over RG-irrelevant short-range contributions, e.g., from the deformation potential. We find that the latter terms generically cause a quasi-particle decay rate scaling as $\sim T^3$. In fact, for insulators and semiconductors,  the piezoelectric interaction is known to dominate small-$q$ scattering if it is allowed by  crystal symmetries \cite{Yu}. We emphasize that the piezoelectric interaction is marginal only in three spatial dimensions. In 2D systems, the corresponding operator is relevant instead.  In practice, such interactions are then screened above a length scale defined
by the bare coupling constant. 

Finally, in view of the symmetry property $e_{ijk}=e_{ikj}$, it is customary to express the piezoelectric tensor in Voigt notation \cite{Mahan},
\be \label{voigt}
e_{ijk}=e_{i(jk)}\mapsto e_{im},
 \quad m=1,\dots,6,
\ee
where matrix elements with $(11)\mapsto 1$, $(22)\mapsto 2$, and $(33)\mapsto 3$ correspond to tension or compression, and those with $(23)=(32)\mapsto 4$, $(13)=(31)\mapsto 5$ and $(12)=(21)\mapsto 6$ describe shear. 
Depending on the crystal symmetry,
the various components in Eq.~\eqref{voigt} may be related to one another or they could vanish identically,  see Ref.~\cite{Nelson} for useful tables. For instance, for TaAs with space group $I4_1md$, No.~109, one finds only three independent components, namely $e_{15}$, $e_{31}$ and $e_{33}$. Their respective values have been computed by \emph{ab initio} methods \cite{Buckeridge}. 

\subsection{Electron-electron interactions}\label{sec2d}

As we show below, the piezoelectric interaction \eqref{piezoqdep} generates a long-range e-e interaction that is attractive in the low-frequency limit where retardation effects can be neglected.  This phonon-mediated  potential has a characteristic angular anisotropy and competes with the repulsive Coulomb interaction in undoped WSMs.  We therefore also include Coulomb interactions from now on.  

\begin{figure}
\begin{centering}
\includegraphics[width=0.8\columnwidth]{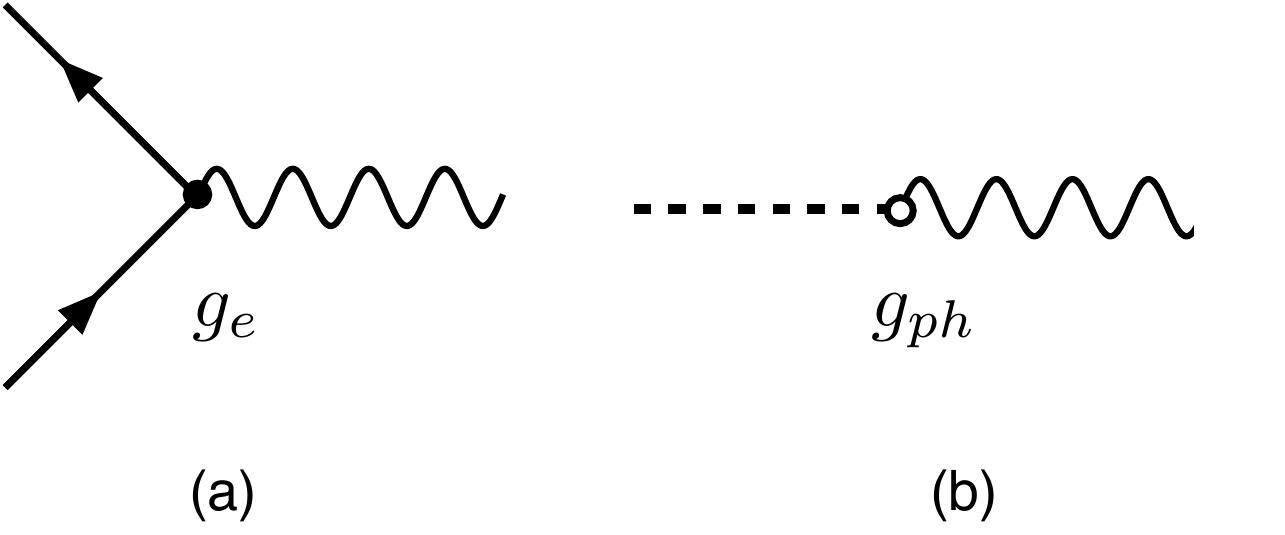}
\par\end{centering}
\caption{\label{fig1} Feynman diagrams for the  vertices in Eq.~\eqref{stotal}, coupling the field $\varphi$ (wiggly curve) to (a) electrons (solid line) and to (b) phonons (dashed). The Coulomb (piezoelectric) vertex $\sim g_e$ ($\sim g_{ph}$) is shown as filled (open) circle.   }
\end{figure}

To that end, we express the Euclidean action of the system in local form by introducing a scalar bosonic Hubbard-Stratonovich field $\varphi(x)$, see Refs.~\cite{Yang,Throckmorton}.  Loosely speaking, the   field $\varphi$ describes photon modes mediating Coulomb interactions. It couples to the sources of the electric field, which include both the conduction electron density and the effective charge density generated by strain via the piezoelectric effect.  With the phonon action $S_{\rm ph}$ in Eq.~\eqref{phononaction}, we  start from the total action
\bea \label{stotal}
S&=&S_{\rm ph}+\int d^4x\Bigl[Z^{-1}_\psi \psi^\ast\partial_\tau \psi -iv \psi^\ast ({\bm\nabla}\cdot\boldsymbol\sigma) \psi \\ \nonumber
&+& \frac{ Z^{-1}_\varphi}{2} ({\bm \nabla}\varphi)^2 + ig_{ e} \psi^\ast\psi\, \varphi  + ig_{ph}  \sum_{jkl} e_{jkl} \partial_j \varphi\, u_{kl}  \Bigr] .
\eea
The bare weight of the fermion (Coulomb) field is given by $Z_\psi=1$ ($Z_\varphi=1$).  These factors could, however, change during the RG flow, see Sec.~\ref{sec3}. 
The partition function is thereby expressed as a functional integral over the fermionic Grassmann fields $(\psi,\psi^*)$,  the displacement field $\bf u$,
and the   field $\varphi$, i.e.,
${\cal Z} = \int {\cal D}[\psi,\psi^\ast, {\bf u}, \varphi] e^{-S}$ \cite{Altland}.
For simplicity, we here assumed isotropic Fermi velocities, cf.~Eq.~\eqref{isotropic}, but we also discuss the general case in Sec.~\ref{sec3}. The action \eqref{stotal} contains two interaction vertices with couplings $g_{e}$
and $g_{ph}$. Their diagrammatic representation is shown in Fig.~\ref{fig1}.

In order to verify that Eq.~\eqref{stotal} makes sense, let us now integrate out the  bosonic field $\varphi$.  With $\rho_e=\psi^\ast \psi$ and switching to Fourier space ($d^4 q=d\omega d^3 \mb q$), the interacting part of the action is then given by
\bea  \label{sint}
S_{\rm int}&=& \int \frac{d^4 q}{(2\pi)^4} \Biggl( \frac{g_e^2}{2|\mb q|^2} \rho_e(q) \rho_e(-q) + \\ 
\nonumber &&\qquad +\, g_eg_{ph} \sum_{ijk} e_{ijk} \frac{q_iq_j}{|\mb q|^2} u_k(q) \rho_e(-q) +  \\ &+& \nonumber
\frac{g_{ph}^2}{2} \sum_{ijk}\sum_{lmn} e_{ijk}e_{lmn} \frac{q_{i}q_{j}q_{l}q_{m}}{|\mb q|^2} u_k(q) u_n(-q)
\Biggr).
\eea
The first term corresponds to the Coulomb e-e interaction upon choosing $g_e^2=e^2/\varepsilon$, while
the second term reproduces the piezoelectric interaction \eqref{piezoqdep} for $g_e g_{ph}=e/\varepsilon$. 
The bare couplings are therefore given by
\be\label{gedef}
g_e = \frac{e}{\sqrt{\varepsilon}},\qquad g_{ph}= \frac{1}{\sqrt{\varepsilon}}.
\ee
We emphasize that the charge $e$ is associated only with the Coulomb vertex $\sim g_e$ in Fig.~\ref{fig1}.
In Eqs.~\eqref{stotal} and \eqref{sint}, 
we have tacitly assumed that intra- and inter-node Coulomb interactions can be taken identical.  Since the effects considered in our paper come from the long-range $1/r$ tail of the Coulomb potential, the couplings between long-wavelength density fluctuations $\rho_h$ and $\rho_{h'}$ of electrons near the Weyl nodes $h$ and $h'$, respectively, are approximately described by the same potential.
The last term in Eq.~\eqref{sint} describes the energy density associated with strain-induced electric fields. Being quadratic in the strain tensor, this contribution generates the so-called piezoelectric stiffening correction, see Ref.~\cite{Nelson} for details. This modification of  the phonon
dispersion typically acts to increase  sound velocities \cite{Nelson,Rinkel}. 
Since here our main interest is centered on electronic instabilities, we will simply assume that the phonon velocities $c_J(\hat {\mb q})$ in Eq.~\eqref{phonondisp} already 
incorporate piezoelectric stiffening to all orders in $g_{ph}$.

\begin{figure}
\begin{centering}
\includegraphics[width=\columnwidth]{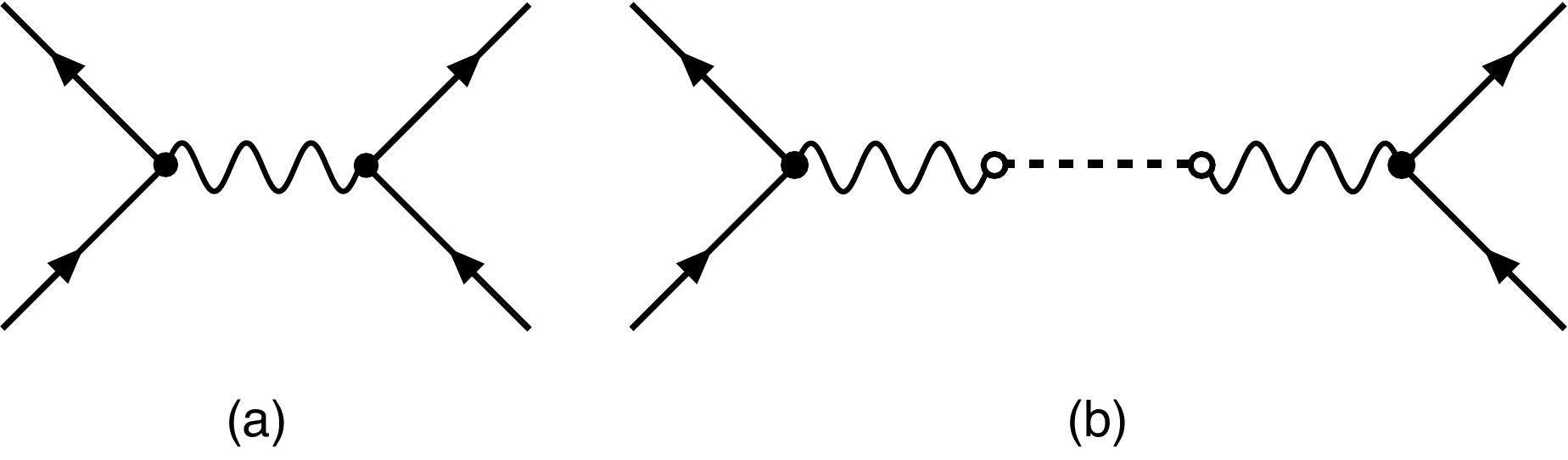}
\par\end{centering}
\caption{\label{fig2}  Effective e-e interaction at tree level. (a) Repulsive Coulomb interaction. (b) Phonon-mediated e-e 
interaction, see Eq.~\eqref{phon1}.  }
\end{figure}

Next we discuss the effective interaction potential between two Weyl fermions described by the above theory.
The two diagrams determining the effective e-e interaction at tree level, i.e., to lowest nontrivial order in perturbation theory, are illustrated in Fig.~\ref{fig2}.
In particular,  Fig.~\ref{fig2}(b) defines a retarded e-e interaction potential, $V_{\rm ph}(q)$, mediated by the piezoelectric interaction, where $q=(i\omega,\mb q)$ is the exchanged four-momentum. Using Eq.~\eqref{stotal} and the phonon propagator in Eq.~\eqref{phonD}, we find 
\be \label{phon1}
V_{\textrm{ph}}(q)=\sum_J\frac{g_e^2 g_{ph}^2/\rho_0}{(i\omega)^2-\Omega^2_J(\mb q)} \sum_{ijk}\frac{\left|e_{ijk} q_iq_j\epsilon^J_k(\hat {\mb q})\right|^2}{|\mb q|^4}.
\ee
Neglecting retardation effects by going to the static limit, $\omega \to 0$, the potential can be written as
\be\label{phononpotential}
V_{\textrm{ph}}(\mb q)=-\frac{g_e^2 g_{ph}^2}{\rho_0 \mb q^2}\gamma(\hat{\mb q}), \quad\gamma(\hat {\mb q})=\sum_{J=1}^3 \gamma^{}_J(\hat{\mb q})
,\ee
with the anisotropy functions
\be\label{functionGJ}
\gamma^{}_J(\hat{\mb q})= \frac{1}{c^2_J(\hat{\mb q}) |{\mb q}|^4}\Biggl|\sum_{ijk}    e_{ijk} q_i q_j   \epsilon^J_k(\hat{\mb q})\Biggr|^2,
\ee
which describe the angular dependence of the phonon-mediated interaction.
We emphasize that $\gamma(\hat{\mb q})>0$ for all directions $\hat{\mb q}$, and thus the   interactions in Eq.~\eqref{phononpotential} are always attractive. 
Combining Eq.~\eqref{phononpotential} with the long-range Coulomb interaction in Fig.~\ref{fig2}(a), we arrive at 
the total e-e interaction potential 
\be\label{vtot}
V_{\textrm{tot}}(\mb q)=\frac{g_e^2}{ \mb q^2}\left(1-\frac{g_{ph}^2}{\rho_0}\gamma(\hat{\mb q})\right).
\ee

Let us now consider WSMs in the $4mm$ crystal class, which in particular includes TaAs, and also use the simplifications in Eqs.~\eqref{isotropic} and Eq.~\eqref{isophon}. 
In Voigt notation, see Sec.~\ref{sec2c}, we define the ratios of  piezoelectric coefficients  
\be \label{AB}
A=\frac{e_{15}}{e_{33}} ,\qquad  B=\frac{e_{31}}{e_{33}} .
\ee
The anisotropy function
$\gamma=\gamma(\theta)$ now depends only on the polar angle $\theta$ of $\hat{\mb q}$.
To evaluate  Eq.~\eqref{functionGJ}, the polarization unit vectors are parametrized  as
\be
\boldsymbol\epsilon^1(\hat{\mb q})=i\hat {\mb q}, \quad
\boldsymbol\epsilon^2(\hat{\mb q})=\frac{i\hat{\mb z}\times \hat {\mb q}}{|\hat{\mb z}\times \hat {\mb q}|},
\quad
\boldsymbol\epsilon^3(\hat{\mb q})=\hat {\mb q}\times \boldsymbol\epsilon^2(\hat{\mb q}),
\ee
leading to
\bea \nonumber
\gamma_1(\theta)&=&\frac{ e_{33}^2}{c_{ph}^2}\cos^2\theta\left[1+(2 A +B-1) \sin^2\theta  \right ]^2,\\
\gamma_2(\theta)&=&0, \label{ff2}\\ \nonumber
\gamma_3(\theta)&=&\frac{ e_{33}^2}{c_{ph}^2} \sin^2\theta\left[\left(B-1\right) \cos^2 \theta +A \cos (2 \theta )\right]^2.
\eea
The contribution from the $J=2$ transverse mode, where the polarization is always perpendicular to $\hat{\mb z}$, vanishes identically. More generally,  $\gamma_J(\theta)=0$ whenever $\boldsymbol \epsilon^J\cdot \hat{\mb z}=0$.

In the simplest approximation, one may just average over the directions $\hat{\mb q}$
in Eq.~\eqref{phononpotential}, see   Refs.~\cite{MahanBook,Mahan}. We write the angular-averaged total interaction
potential as
\be\label{vtotav}
\bar V_{\textrm{tot}}(\mb q)=\frac{g_e^2(1-\bar \gamma)}{\mb q^2}.
\ee
For the 4$mm$ crystal class, we find from Eq.~\eqref{ff2}  
\be\label{bargamma}
\bar\gamma=\frac{g_{ph}^2}{2\rho_0} \int_0^\pi d\theta \sin(\theta)\gamma(\theta)
=   \frac{w_\gamma}{\rho_0} \left(\frac{g_{ph} e_{33}}{c_{ph}}\right)^2,
\ee
with the coefficient 
\be \label{CPi}
w_\gamma=\frac{1}{15}\left[10 A^2+4 A (B+1)+2 B^2+3\right].
\ee
Clearly, for $\bar \gamma>1$, the averaged total interaction \eqref{vtotav} is attractive. One thus expects a gapped  superconducting phase with $s$-wave singlet pairing.  However, as we show in Sec.~\ref{sec4}, for $\bar\gamma<1$, one may also encounter more exotic superconducting phases exhibiting, e.g., nodal-line triplet pairing.

\subsection{Parameter estimates}\label{sec4a}

 \begin{figure}
\begin{centering}
\includegraphics[width=\columnwidth]{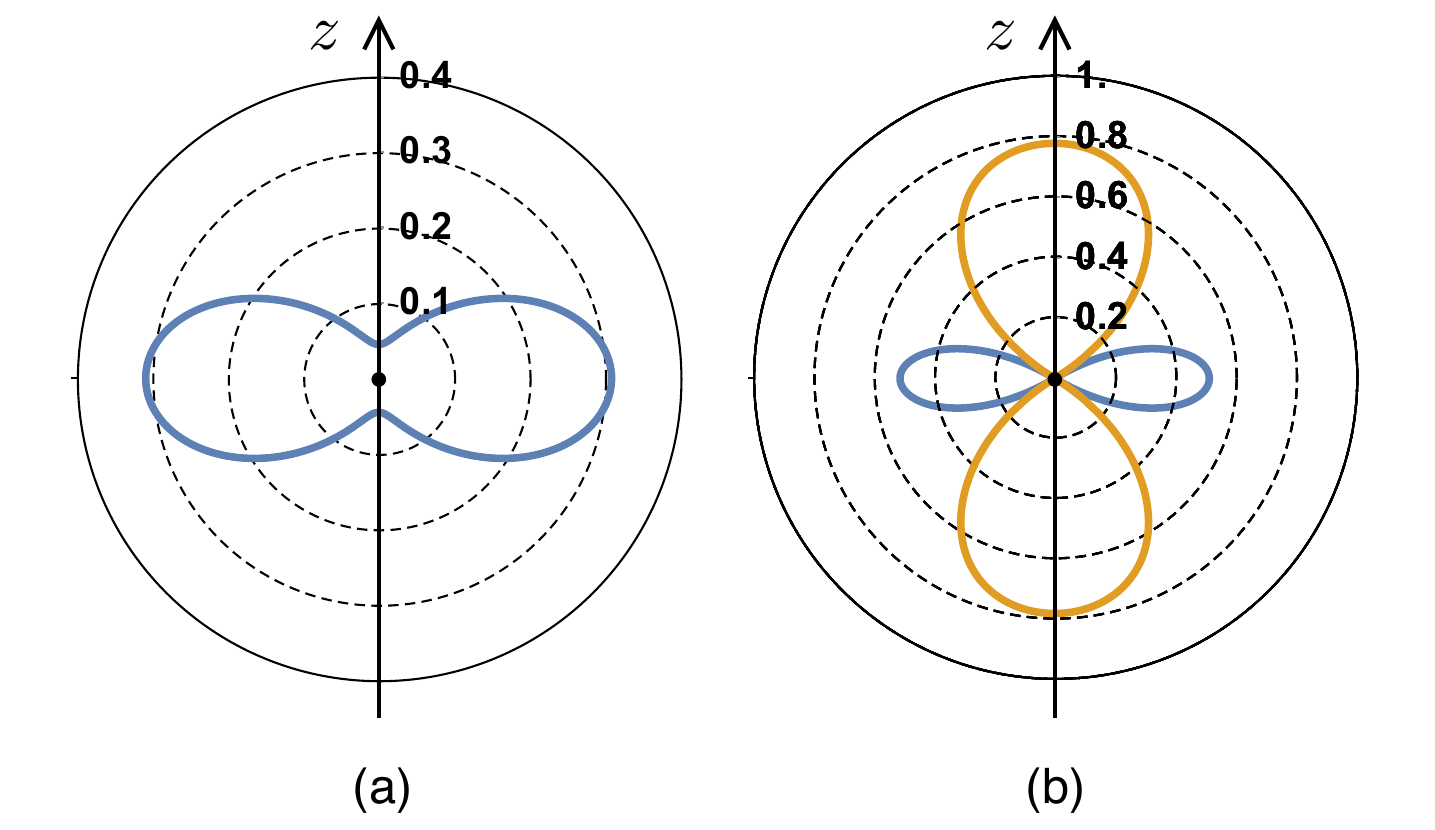}
\par\end{centering}
\caption{\label{fig3} (a) Polar plot of the anisotropy function $\gamma(\theta)$ in  Eq.~\eqref{phononpotential} for the case of TaAs, with $\gamma(\theta)$  multiplied by $g_{ph}^2/\rho_0$. We take
$\varepsilon=20\varepsilon_0$, where $\varepsilon_0$ is the free-space permittivity. The piezoelectric tensor values are taken from Ref.~\cite{Buckeridge}, where we get 
$\bar\gamma\simeq 0.20$ in Eq.~\eqref{bargamma}.  The blue color indicates that the phonon-mediated interaction   is always attractive.
(b) Effective anisotropy function of the total e-e interaction potential in Eq.~\eqref{vtot}, where we adjust $e_{33}$ such that $\bar\gamma=0.97$.
Blue again indicates attraction while orange represents repulsion. }
\end{figure}

To get concrete predictions from our theory, we need information about the  piezoelectric coefficients \cite{Xue,Berlincourt,Acosta},  the permittivity $\varepsilon$, the mass density $\rho_0$, and the Fermi as well as the sound velocities. Since  in TaAs 
the lattice parameters are $a_\perp\simeq 3.43${\AA} and $a_3\simeq 11.6${\AA}, and the conventional unit cell contains 4 Ta and 4 As ions, 
the  mass density is  $\rho_0\simeq 1.24 \times 10^4$ kg/m$^3$. For simplicity,  we here adopt the simplifying assumptions in Eqs.~\eqref{isotropic} and \eqref{isophon}. For the Fermi velocity, we take $\hbar v\simeq 2$~eV{\AA} \cite{Huang}, which corresponds to $v\simeq 3 \times 10^5$ m/s.  The sound velocity  is assumed to be given by $c_{\rm ph}\simeq 6\times 10^3$~m/s, cf.~the value quoted in Ref.~\cite{Guo} for TaN. For the piezoelectric tensor of TaAs \cite{Buckeridge}, we use $e_{33}=-1.89 $ Cm$^{-2}$ and the ratios in   Eq.~\eqref{AB} are $A \simeq -2.62$ and   $ B\simeq -0.43$. This gives $w_\gamma\simeq4.40$. 
  Using the rough estimate $\varepsilon\approx  20 \varepsilon_0$ \cite{Throckmorton}, we obtain 
$\alpha_{\rm eff}\approx 0.24$ and $\bar \gamma\approx 0.20$. The latter is well below the critical value $\bar \gamma=1$. However, the value of $\bar\gamma$ could in principle be   higher in other materials which might have, for instance, larger piezoelectric coefficients or smaller permittivity. 

Moreover, the approximation in Eq. \eqref{vtotav} neglects the angular anisotropy of the effective interaction.  A polar plot of $\gamma(\theta)$ based on our estimates for TaAs is shown in Fig.~\ref{fig3}(a). The attractive interaction strength is maximal for $\theta=\pi/2$. This shape of $\gamma(\theta)$ is representative of the regime $|e_{15}|>|e_{33}|>|e_{31}|$, which is also realized for the paradigmatic piezoelectric insulator BaTiO$_3$ \cite{Xue}. For TaAs, the total e-e interaction potential is repulsive in all directions. However, for higher values of $\bar\gamma$ and depending on the relative strength of the Coulomb and the piezoelectric terms,   there may be  directions along which the total interaction potential becomes attractive even for $\bar\gamma<1$. In this case,  superconducting phases could   be possible despite the effective repulsion in the $s$-wave channel. In Fig.~\ref{fig3}(b), we show the angular dependence of the total e-e interaction potential  \eqref{vtot} for  $\bar \gamma=0.97$. In this case, the total e-e interaction potential     changes sign as a function of $\theta$ and becomes attractive for $\theta\simeq \pi/2$.

\section{RG analysis} \label{sec3}

In this section, we turn to the derivation and solution of the one-loop RG equations.  
In an infinitesimal RG  step, the flow parameter changes as $\ell\mapsto \ell+d\ell$,
where $\Lambda(\ell)=e^{-\ell}\Lambda_0$ is the running high-energy bandwidth cutoff with  bare value $\Lambda_0$. 
We obtain the RG equations by the standard momentum-shell integration approach, where in each RG step
 one integrates over all field modes appearing in the partition function with energies in the shell $\Lambda(\ell+d\ell)< E<\Lambda(\ell)$. 
 The resulting contributions to the partition function are then taken into account by  renormalization of the various couplings in the action, see Refs.~\cite{Cardy,Altland}.

\begin{figure}
\begin{centering}
\includegraphics[width=0.8\columnwidth]{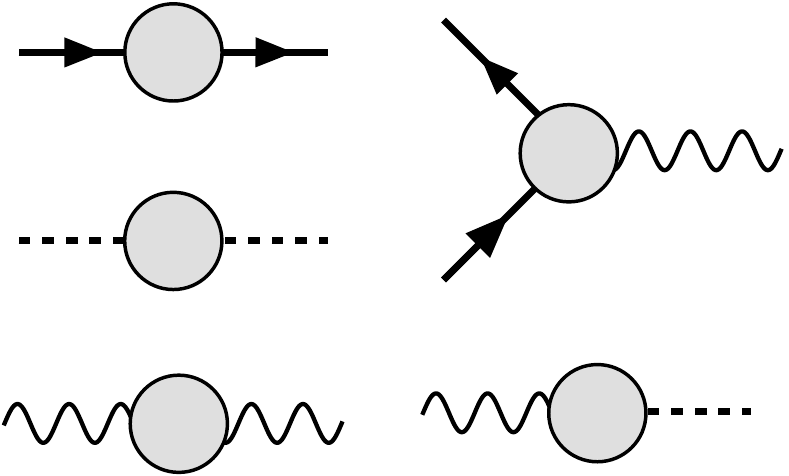}
\par\end{centering}
\caption{\label{fig4} Schematic form of the possible amplitudes generated by the local field theory in Eq.~\eqref{stotal},
where shaded regions represent dressed vertices in a perturbative expansion.  }
\end{figure}

 We start from the observation that for the local field theory \eqref{stotal},  perturbative expansions of physical observables  involve only  diagrams of the types shown in Fig.~\ref{fig4}. In all these   diagrams, 
fermion loop contributions always involve   the Coulomb vertex $\sim g_e$. This fact can be rationalized  by recalling that the piezoelectric interaction also arises from an expansion of the Coulomb potential, see Sec.~\ref{sec2c}. The vertex $g_{ph}$ only appears in regular, perturbative corrections to the Coulomb propagator. 
 At the one-loop level, perturbation theory in $g_e$ generates the diagrams in Figs.~\ref{fig5}(a), \ref{fig5}(b) and \ref{fig5}(c), which are precisely the diagrams that govern the one-loop renormalization of e-e interactions in the absence of phonons \cite{Throckmorton}. 

Within the static approximation with the angular-averaged interaction potential in Eq. \eqref{vtotav},   the piezoelectric interaction is combined with the Coulomb e-e interaction and its effect amounts to replacing $g_e^2\mapsto g_e^2(1-\bar\gamma)$. As a consequence, 
 the essential physics of the system can be studied in terms of a single dimensionless coupling, namely
the effective fine structure constant
\be\label{finestructure}
\alpha_{\textrm{eff}}=\frac{g_e^2(1-\bar \gamma)}{4\pi v}. 
\ee
Within this static approximation, the RG equation for $\alpha_{\textrm{eff}}$ at the one-loop level follows from the diagrams in  Figs.~\ref{fig5}(a), \ref{fig5}(b) and \ref{fig5}(c).  The result is    \cite{Throckmorton}
\be\label{aflow}
\frac{d\alpha_{\textrm{eff}}}{d\ell}=-\frac{2(N+1)}{3\pi}\alpha_{\textrm{eff}}^2. 
\ee 
Therefore,  the system  flows to strong coupling when the effective fine structure constant becomes negative. This happens for sufficiently strong piezoelectric coupling, in the regime $\bar\gamma>1$.  

The strong-coupling phase realized for $\bar \gamma>1$ is expected to be an intrinsic superconductor since the attractive e-ph interaction then dominates over the repulsive Coulomb interaction.  Previous work  \cite{Cho,Hosur2013,Bednik} has discussed intrinsic superconductivity in \emph{doped}  WSMs.  
The new element in our system is the long-range  e-e interaction resulting  from a combination of unscreened Coulomb and  piezoelectric interactions. 
We recall that the standard BCS formula for the superconducting gap is given by $\Delta\sim e^{-1/\nu_F |\lambda|}$, where $\nu_F$ is the normal density of states at the Fermi level and  $\lambda$ denotes the strength of the short-range attractive interaction. For vanishing $\nu_F$, intrinsic superconductivity is not possible unless the short-range interaction exceeds a critical coupling  of the order of the electronic bandwidth, far beyond the perturbatively accessible regime. As we will see in Sec.~\ref{sec4}, the long-range character of the piezoelectric interaction allows for the opening of a finite gap even for the undoped case  with  $\nu_F=0$. In this case, the gap is a function of the dimensionless parameter $\alpha_{\textrm{eff}}<0$.  Eliminating the need for doping to realize superconductivity in WSMs is important because the density of states cannot be made very large if one wants to stay below the energy scale $vb$, where $b$ is the momentum separation between two  Weyl nodes. In fact, at high energies, nonlinearities will appear in the dispersion relation. 
 
\subsection{RG equations beyond the static approximation}\label{sec3a}

We can use the RG approach to analyze how the piezoelectric interaction affects the running couplings in the effective action \eqref{stotal} beyond the static approximation, i.e., including retardation effects. After   performing an infinitesimal RG transformation and rescaling   $\psi\mapsto (1+\delta Z_\psi/Z_\psi)^{1/2}\psi$ and $\varphi\mapsto (1+\delta Z_\varphi/Z_\varphi)^{1/2}\varphi$ to absorb the field renormalizations, we obtain a correction to the effective action of the form\bea
\delta S&=&\int d^4x\Bigl[  -iv \left(1+\frac{\delta v}{v}+\frac{\delta Z_\psi}{Z_\psi}\right)\psi^\ast ({\bm\nabla}\cdot\boldsymbol\sigma) \psi \nonumber\\ 
&& + ig_{ e}\left(1+\frac{\delta g_e}{g_e}+\frac{\delta Z_\psi}{Z_\psi}+\frac12\frac{\delta Z_\varphi}{Z_\varphi}\right) \psi^\ast\psi\, \varphi  \\
&&+ ig_{ph} \left(1+\frac{\delta g_{ph}}{g_{ph}}+\frac12\frac{\delta Z_\varphi}{Z_\varphi}\right) \sum_{jkl} e_{jkl} \partial_j \varphi\, u_{kl}  \Bigr] .\nonumber
\eea
We can compute $\delta g_e$ and $\delta g_{ph}$ from the corresponding vertex corrections, whereas $\delta Z_\psi$ and $\delta v$ stem from the electron self-energy and $\delta Z_\varphi$ from the polarization insertion in the Coulomb propagator. The corrections can then be absorbed as a renormalization of the parameters $v$, $g_e$ and $g_{ph}$.

\begin{figure}
\begin{centering}
\includegraphics[width=.9\columnwidth]{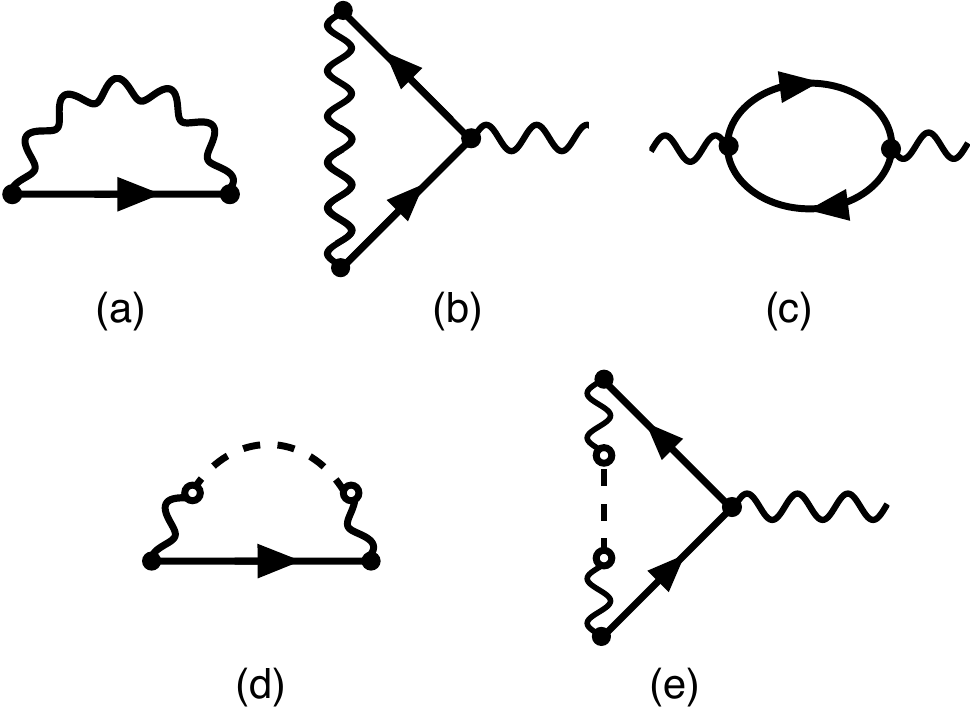}
\par\end{centering}
\caption{\label{fig5} Diagrams contributing to the one-loop RG equations. 
(a) Coulomb correction to the electronic self-energy.
(b) Vertex correction due to Coulomb interaction. 
(c) Polarization bubble inserted in the Coulomb propagator.
(d) Piezoelectric correction to the electronic self-energy.
(e) Piezoelectric vertex correction.}
\end{figure}

At the one-loop level and at lowest order in $g_{ph}$, the  contributions from the piezoelectric interaction are represented by  the  diagrams shown in    Figs.~\ref{fig5}(d) and  \ref{fig5}(e). The latter  are generated by taking into account the (non-divergent) correction to  the  Coulomb  propagator at order $g_{ph}^2$. In the following we will now separately discuss each of the  five diagrams in Fig.~\ref{fig5}.  

\subsubsection{Coulomb correction to the electronic self-energy}\label{sec3a1}

The standard rainbow diagram in Fig.~\ref{fig5}(a) describes the lowest-order  correction to the electronic self-energy due to Coulomb interactions.
A well-known consequence of this contribution is a renormalization of the Fermi velocities.
Related effects have been predicted and experimentally observed for graphene \cite{Elias}. 
The diagram in Fig.~\ref{fig5}(a) yields the self-energy
\be\label{selfenergy1}
\Sigma_{ee}(p)=-g_e^2\int\frac{d^4q}{(2\pi)^4}\frac{1}{ \mb q^2}\mathbb G( p+q),
\ee
with $p=(i\omega,\mb p)$.  We evaluate Eq.~\eqref{selfenergy1} in App.~\ref{app1}, where we show that $\Sigma_{ee}$ does not depend on the frequency $\omega$ and hence no field renormalization arises from this term, $\delta Z_\psi=0$.
Integrating out the modes of the field $\varphi$  within the high-energy momentum shell  and keeping only self-energy terms linear in the momentum $\mb p$,  we arrive at the self-energy correction 
\be\label{selfenergy2}
\delta\Sigma_{ee}(\mb p)= \frac{g_e^2}{8\pi^2}  \left(\eta_\perp \mb p_\perp\cdot \boldsymbol\sigma_\perp+\eta_3  p_3\sigma_3
\right)d\ell,
\ee
where the numbers $\eta_\perp$ and $\eta_3$ depend on the Fermi velocity ratio $v_3/v_\perp$, cf.~App.~\ref{app1}. By comparing with Eq.~\eqref{Hplus}, we see that Eq.~\eqref{selfenergy2} generates a correction to the Fermi velocities $v_\perp$ and $v_3$.
For the isotropic case \eqref{isotropic}, we get $\eta_\perp=\eta_3=4/3$. In this case, we obtain
\be\label{vrenormCoulomb}
\delta v = \frac{g_e^2}{6\pi^2}d\ell  .
\ee 
By itself, this term makes the Fermi velocity increase under the RG flow. 

\subsubsection{Vertex correction due to Coulomb interaction}
\label{sec3a2}
 
Next we turn to the diagram in Fig.~\ref{fig5}(b), which provides a vertex correction due to the Coulomb interaction, 
corresponding to a charge renormalization \cite{Altland}. 
However,  this diagram   actually gives no contribution at all.
In fact, the instantaneous Coulomb interaction does not give rise to charge renormalization for Weyl (or Dirac) fermions at the one-loop level \cite{Kotov}. 
For the corresponding 2D graphene case, charge renormalization is absent also at the two-loop level 
\cite{Kotov}. 

\subsubsection{Coulomb propagator: Polarization bubble}\label{sec3a3}

 At the one-loop level, the self-energy of the    field  $\varphi$ comes from the standard polarization bubble in Fig.~\ref{fig5}(c). 
 Following the analysis of Ref.~\cite{Throckmorton}, the self-energy correction can be absorbed by the field renormalization of $\varphi$,
\be \label{fieldren}
\delta Z_\varphi=-\frac{N g_e^2 }{6\pi^2 v} Z_\varphi d\ell ,
\ee
where  the presence of a fermion loop in the diagram implies that this correction is proportional to the number of Weyl nodes, $2N$. 
For simplicity, we have again assumed isotropic Fermi velocities, see Eq.~\eqref{isotropic}. 

\subsubsection{Piezoelectric self-energy correction}\label{sec3a4}

Next we turn to the electronic self-energy $\Sigma_{ep}(i\omega,\mb p)$ due to e-ph interactions, which to one-loop order
comes from the diagram in Fig.~\ref{fig5}(d). We evaluate this term in App.~\ref{app2}, see Eq.~\eqref{B4}.  
 A non-universal contribution arises for $\omega=\mb p=0$ which can be absorbed by renormalization of the chemical potential.  A similar contribution also comes from e-e interactions, see App.~\ref{app1}, and we eventually require the renormalized chemical potential to be located at the Weyl node.   As discussed in App.~\ref{app2}, for $4mm$ crystal symmetry and again
 using Eqs.~\eqref{isotropic} and \eqref{isophon},  the self-energy correction after momentum-shell integration is given by 
\bea\label{selfenergy3}
\delta\Sigma_{ep}(p)&=&-\frac{1}{4\pi \rho_0}\left(\frac{g_eg_{ph} e_{33}}{c_{ph}}\right)^2 \frac{c_{ph}}{v}\\ \nonumber &\times& \left(\frac{i\omega C_0}{v} \sigma_0 + C_\perp\mb p_\perp\cdot\boldsymbol\sigma_\perp+C_3p_3\sigma_3 \right) d\ell,
\eea
with the numbers $C_0\simeq 1.40$, $C_\perp\simeq 0.29$ and $C_3\simeq 0.83$ for TaAs.  The smallness of the factor $c_{ph}/v\ll 1$, together with the fact that in practice we have $\bar\gamma\lesssim 1$ in Eq.~\eqref{bargamma},
implies that contributions from Eq.~\eqref{selfenergy3} to  RG equations are rather small.  

In marked contrast to the Coulomb case,  we now encounter in Eq.~\eqref{selfenergy3} a term $\Sigma_{ep}\sim \omega$ responsible for field renormalization,
\be\label{fieldren1}
\delta  Z_\psi =-\frac{C_0}{4\pi \rho_0  v} \frac{c_{ph}}{v}
\left(\frac{g_e g_{ph} e_{33}}{c_{ph} }\right)^2 Z_\psi d\ell ,
\ee
implying that the quasi-particle weight $Z_\psi$  decreases under the RG flow.

The $\mb p\ne 0$ terms in Eq.~\eqref{selfenergy3} can be absorbed by renormalization of the Fermi velocities. 
In general, even for initially isotropic velocities, the fact that $C_\perp\ne C_3$ implies that piezoelectric couplings intrinsically generate anisotropic Fermi velocities.
Because we have $c_{ph}/v\ll 1$, however, this Fermi velocity renormalization is typically subleading
against the dominant Coulomb term in Eq.~\eqref{vrenormCoulomb}. For simplicity, we here neglect the RG-generated anisotropy of the Fermi velocities and only focus on the mean value of the Fermi velocity defined as $v=(2v_\perp+v_3)/3$, cf.~Eq.~\eqref{isotropic}. Taking into account Eq.~\eqref{vrenormCoulomb} and using the number $\bar C=(2C_\perp+C_3)/3$, with $\bar C\simeq 0.47$ for TaAs, we obtain another correction to the Fermi velocity 
which must be added to Eq.~\eqref{vrenormCoulomb},
\be\label{vrenormCoulomb2}
\delta v'  =-\frac{g_e^2}{4\pi}    \frac{\bar C c_{ph}}{\rho_0 v} 
\left(\frac{g_{ph} e_{33}}{c_{ph}}\right)^2   d\ell.
\ee 
Since $\bar C>0$, the piezoelectric corrections tend to decrease the Fermi velocities.

 \subsubsection{Piezoelectric vertex correction}\label{sec3a5}

One-loop vertex corrections do arise from the piezoelectric coupling, see the diagram in Fig.~\ref{fig5}(e). This diagram is studied in detail in App.~\ref{app3}.  We obtain a charge renormalization corresponding to the RG flow of the coupling $g_e$ in Eq.~\eqref{gedef}.
For the $4mm$ crystal class, and using again Eqs.~\eqref{isotropic} and \eqref{isophon}, we obtain 
\be\label{deltacharge}
\delta g_e =\frac{C_0}{4\pi\rho_0}\frac{c_{ph}}{v}  \left( \frac{ g_e g_{ph} e_{33}}{c_{ph}} \right)^2  g_e d\ell,
\ee
with $C_0\simeq 1.40$ for TaAs. Note the factor of $c_{ph}/v\ll1$, which is a manifestation of Migdal's theorem for WSMs \cite{Roy2014}. The fact that the same coefficient  $C_0$ governs both the vertex  correction and the field renormalization, see Eq.~\eqref{fieldren1}, 
is due to a Ward identity for electron-phonon interactions \cite{Engelsberg}. 
 We also have  $\delta g_{ph}=0$ because there are no loop corrections in  this vertex.

\subsection{RG equations}\label{sec3b}

\begin{figure}
\begin{centering}
\includegraphics[width=.9\columnwidth]{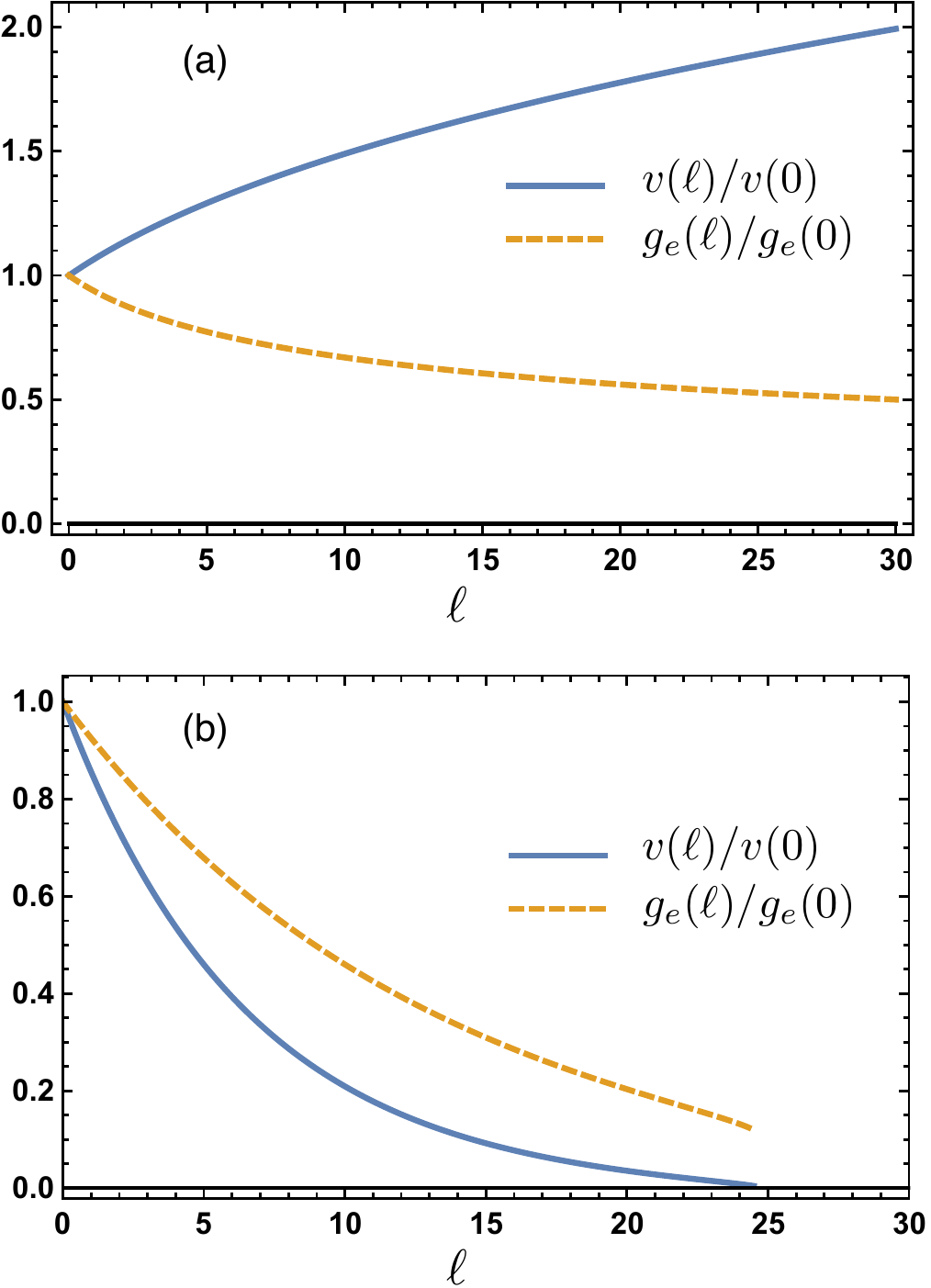}
\par\end{centering}
\caption{\label{figZ} Renormalized Fermi velocity $v(\ell)$ and   Coulomb coupling $g_e(\ell)$ as   functions of the RG flow parameter $\ell=\ln(\Lambda_0/\Lambda)$. (a) Flow diagram obtained using the estimated parameters for TaAs, corresponding to $\bar\gamma=0.20$, but considering  $2N=4$ Weyl nodes.  (b) Flow diagram obtained by enhancing the piezoelectric coefficient $e_{33}$ to reach $\bar \gamma\simeq 75$. Here we stop the RG flow at the scale where the Fermi velocity vanishes, at which point   the WSM becomes unstable.}
\end{figure}

We now collect the results of Sec.~\ref{sec3a}.  The one-loop RG equations are then given by 
\bea \nonumber
\frac{dZ_\psi}{d\ell}&=&-C_0\frac{c_{ph}}{v} \frac{g_e^2}{4\pi v}\frac{g_{ph}^2e_{33}^2}{\rho_0c_{ph}^2}    Z_\psi ,\\ \nonumber
\frac{d Z_\varphi}{d\ell}&=&-\frac{Ng_e^2}{6\pi^2 v} Z_\varphi,\\  
\frac{dv}{d\ell}&=&
\frac{g_e^2}{6\pi^2} \left[ 1 - \frac{3\pi (C_0+\bar C)   }{2} \frac{c_{ph}}{v}\frac{g_{ph}^2e_{33}^2}{\rho_0c_{ph}^2}   \right]\label{RGfull}
,\\ \nonumber
\frac{d g_e}{d\ell}&=&-\frac{Ng_e^3}{12\pi^2 v},\\
\nonumber
\frac{d g_{ph}}{d\ell}&=&-\frac{Ng_e^2 g_{ph}}{12\pi^2 v}.
\eea
We note that on effective length scales beyond the mean free path, disorder effects could modify the above RG equations.
For $g_{ph}=0$, we recover the RG equations  in the absence of phonons, in which case the Coulomb vertex $g_e$ is marginally irrelevant and the Fermi velocity increases monotonically as we lower the energy scale. For $g_{ph}\neq0$, the vertex correction     $\delta g_e/g_e$ due to the piezoelectric interaction gets canceled by the field renormalization $\delta Z_\psi/Z_\psi$ and  $g_e$ still decreases with the RG flow. Solving the RG equations numerically with the initial condition set by the parameters for TaAs, we obtain the flow diagram in Fig. \ref{figZ}(a). 

However, we find that an instability can  arise if the    piezoelectric interaction is strong enough to reverse the flow of the Fermi velocity and make it vanish (or become of the order of the phonon velocity) at some finite energy scale. 
A rough estimate of the condition for this instability is obtained by imposing that $dv/d\ell$  must be negative at the beginning of the RG flow. This requires $\bar\gamma>\frac{2 w_\gamma}{3\pi (C_0+\bar C)}\frac{v}{c_{ph}}$.  While  $C_0$, $\bar C$ and $w_\gamma$  are constants of order unity, the factor of velocity ratio $v/c_{ph}\gg1$ pushes the critical  $\bar\gamma$   to a higher value than estimated within the static approximation. Integrating the RG equations numerically, we find that the renormalized velocity does vanish when we enhance the piezoelectric coefficient such that $\bar\gamma \simeq 75$, as shown in Fig.  \ref{figZ}(b). Therefore, this RG analysis suggests that retardation effects make the WSM phase more stable against a superconducting transition.

\section{Phase diagram and superconductivity}\label{sec4}

We next perform a self-consistent mean-field analysis to locate superconducting regions in the phase diagram within  the static approximation for the total interaction. We   develop the mean-field approach in Sec.~\ref{sec4b} and study the stability of superconducting phases with singlet or triplet pairing.  For  small $\bar \gamma$, the WSM phase remains stable but will be characterized by a sizeable quasi-particle decay rate $\Gamma$.  We determine the dependence of $\Gamma$ on temperature and on the energy of the quasi-particle in Sec.~\ref{sec4c}.

\subsection{Mean field theory}\label{sec4b}

Since pairing involves time-reversed partner states,   we consider the effective inter-node e-e interaction potential $V_{\rm tot}(\mb q)$ in Eq.~\eqref{vtot} for a pair of nodes ($h=1,2$) that are linked by time reversal. The Hamiltonian is then given by
\bea\label{heff1}
H_{\textrm{eff}}&=&\sum_{h=1}^2 \sum_{\mb p}\psi^\dagger_h(\mb p) \left(v \mb p\cdot\boldsymbol\sigma\right) \psi^{\phantom\dagger}_h(\mb p)
\\ \nonumber
&+&\frac1{V}\sum_{\mb k,\mb p,\mb q}V_{\textrm{tot}}(\mb q)\psi^\dagger_1(\mb p+\mb q)\psi^{\phantom\dagger}_1(\mb p)\psi^\dagger_2(\mb k-\mb q)\psi^{\phantom\dagger}_2(\mb k).
\eea
We assume the static approximation for the total e-e interaction, as done in the standard BCS theory for the normal-metal-superconductor transition. 
 While phonon-induced retardation effects could be included within Eliashberg theory, we here explore only the static case defined by Eq.~\eqref{heff1}. 
We expect to encounter a superconducting phase for $\bar \gamma>1$, see Eq.~\eqref{bargamma}, where the effective interaction $V_{\rm tot}$ will be attractive in all directions and the order parameter should describe $s$-wave singlet pairing.  
However, it is worth mentioning that  the   breaking  of spin-rotational invariance by spin-orbit coupling in WSMs blurs the distinction between singlet and triplet  pairing   \cite{Cho}. In fact, a mixing of singlet and triplet components is generic for non-centrosymmetric superconductors \cite{Sigrist,Yip}. 
With this caveat in mind, we now implement the mean-field approximation for $H_{\rm eff}$ in Eq.~\eqref{heff1}. 

We consider a generic spin-matrix order parameter, $\boldsymbol\Xi(\mb k)$, defined by 
\be\label{BigXi}
\left\langle\psi_{1\sigma}(\mb k)\psi_{2\sigma'}(-\mb k+\mb q)\right\rangle=\delta_{\mb q,0} \, \left[{\bm\Xi}(\mb k)i\sigma_2\right]_{\sigma\sigma'}.
\ee 
The gap function then also corresponds to a complex-valued spin matrix,
\be\label{DeltaXi}
{\bm\Delta}(\mb p)=-\frac1V\sum_{\mb k}V_{\textrm{tot}}(\mb p-\mb k){\bm\Xi}(\mb k).
\ee
Using four-component Nambu spinor operators \cite{MahanBook},
\be
\Psi(\mb p)=\left(\begin{array}{c}\psi_1(\mb p)\\
i\sigma_2\psi^\dagger_2(-\mb p)\end{array}\right), \quad
\psi_h(\mb p)= \left(\begin{array}{c}
\psi_{h,\uparrow}(\mb p)\\
\psi_{h,\downarrow}(\mb p)\end{array}\right),
\label{Nambu}
\ee
the standard mean-field decoupling scheme yields the Bogoliubov-de-Gennes (BdG) Hamiltonian
\bea\nonumber
H_{\textrm{BdG}}&=&
\sum_{\mb p}
\left(\Psi^{\dagger}(\mb p) {\cal H}_{\rm BdG}(\mb p)
\Psi(\mb p)
+\textrm{Tr}\left[{\bm\Delta}^\dagger(\mb p){\bm\Xi}(\mb p)\right]\right),\\ \label{BdG} 
&& {\cal H}_{\rm BdG}(\mb p) =
\left(\begin{array}{cc}
v\boldsymbol \sigma\cdot\mb p & {\bm\Delta}(\mb p)\\
{\bm\Delta}^\dagger(\mb p)&-v\boldsymbol \sigma\cdot\mb p
\end{array}\right).
\eea
We will now examine the conditions for superconducting phases with singlet vs triplet pairing. 

\subsubsection{Singlet  pairing}\label{sec4b1}

For the case of singlet pairing, we write ${\bm\Delta}(\mb p)=\Delta_0(\mb p)\sigma_0$ in a gauge where the scalar function $\Delta_0(\mb p)$ is real valued. 
Diagonalizing ${\cal H}_{\rm BdG}(\mb p)$ in Eq.~\eqref{BdG}, one finds the eigenvalues $\pm E_s(\mb p)$
with $E_s(\mb p)=\sqrt{v^2 {\bf p}^2+\Delta_0^2(\mb p)}$.
 The gap equation then follows from Eq.~\eqref{DeltaXi} by noting that Eq.~(\ref{BigXi}) is solved by a spin-isotropic matrix, ${\bm \Xi}(\mb k)=
 \frac{\Delta_0(\mb k)}{2E_s(\mb k)}\sigma_0$.  Using the averaged interaction potential in Eq.~\eqref{vtotav} with $\bar \gamma$ in Eq.~\eqref{bargamma},
the solution follows by assuming a constant gap function,  $\Delta_0(\mb k)=\Delta_0$, corresponding to $s$-wave pairing.  For $\Delta_0\ne 0$, with Eq.~\eqref{finestructure} we arrive at the gap equation
 \bea\nonumber
1&=&-\frac{1-\bar \gamma}{4\pi^2}\int_0^{b}dk\,k^2\frac{g_e^2}{ k^2\sqrt{v^2k^2+\Delta_0^2}}
\\&=&
-\frac{\alpha_{\rm eff}}{\pi}\ln\left(\frac{2vb}{\Delta_0}\right),\label{solutiongap}
\eea
where the large-momentum cutoff $b$ corresponds to the momentum separation between different Weyl nodes. 
For $\alpha_{\rm eff}<0$, corresponding to $\bar  \gamma>1$, we then find the isotropic gap
\be\label{delta0singlet}
\Delta_0=2vb\, e^{-\pi/ |\alpha_{\rm eff}|}.
\ee
Assuming that $\Delta_0$ has the same sign at both Weyl nodes \cite{Hosur2013,Meng},
we obtain a topologically trivial gapped superconductor with conventional $s$-wave singlet pairing. However, it is worth noting again that a finite gap emerges  even though $\nu_F$ vanishes at the Fermi level. Technically,  the $1/{\bf k}^2$ momentum dependence of the long-range interaction potential compensates the  density-of-states factor $k^2$ in Eq.~(\ref{solutiongap}). 

\subsubsection{Nodal-line triplet pairing}
\label{sec4b2}

We next investigate the possibility of other superconducting phases at $\bar\gamma<1$, where the effective interaction strength is repulsive along certain directions but a significant attractive component exists  near the $q_3=0$ plane, see Fig.~\ref{fig3}(b).
A general superconducting order parameter can be written as
\be \label{ordpar}
{\mb\Delta}(\mb k)=\Delta_0(\mb k)\sigma_0+\mb a(\mb k)\cdot \boldsymbol \sigma,
\ee
where $\Delta_0(\mb k)$ is a real scalar function  and $\mb a(\mb k)$ is a complex vector field.  For ${\mb a}\ne 0$, the superconducting phase has a triplet pairing component \cite{Cho}.
We require that the BdG Hamiltonian \eqref{BdG} preserves time-reversal symmetry, which implies  the conditions
\be
\Delta_0(-\mb k)=\Delta_0(\mb k),\quad \mb a^*(-\mb k)=-\mb a(\mb k).\label{cond1}
\ee
We then expand Eq.~\eqref{ordpar} to first order in $\mb k$,  where time-reversal symmetry and
 Eq.~\eqref{cond1} imply  
\be\label{mftpar}
\Delta_0(\mb k)=\Delta_0 ,\quad
{\mb a}({\mb k})= {\bm M}\cdot {\mb k} + i {\mb a}_2 .
\ee
Here ${\bm M}$ is a real $3\times 3$ matrix and  
the vector ${\mb a}_2$ also has real entries.

Next, in order to reduce the number of mean-field parameters, we take into account global spin and orbital rotation symmetry around the $z$-axis for tetragonal crystal symmetry. 
In this argument, we assume that these symmetries are approximately realized even when expanding around the Weyl nodes. This approximation becomes exact if the Weyl points are separated along the $z$-axis in momentum space.
Indeed,  a state that minimizes the energy should take advantage of the anisotropy in the effective interaction \eqref{vtot}. We thus take $\mb a_2=a_2\hat {\mb z}$ and ${\bm M}={\rm diag}(a_{\perp},a_{\perp},a_{\parallel})$, leaving us with only
 four mean-field parameters in Eq.~\eqref{mftpar}. For $\Delta_0=0$, the eigenvalues of ${\cal H}_{\rm BdG}(\mb k)$ are
given by $\pm E_t(\mb k)$ with
\bea\label{spectrumnodal}
E^2_t(\mb k)&=& v^2 \mb k ^2+a_{\perp}^2 \mb k_\perp^2+a_{\parallel}^2 k_3^2+a_2^2 \\ \nonumber
&\pm& 2 |\mb k_\perp|\sqrt{(v^2+ a_{\perp}^2) a_2^2+v^2 k_3^2( a_{\perp}-a_{\parallel})^2} .
\eea 
For $a_2=0$, the energy only vanishes at $\mb k=0$, and  each of the original Weyl nodes  splits into  two   Bogoliubov-Weyl nodes, similar to  the result of  Ref.~\cite{Meng} for pairing between nodes with opposite chirality. 
For $a_2\neq0$, the spectrum instead exhibits a \emph{nodal ring} in the $k_3=0$ plane,
\be
|\mb k_\perp|=\frac{|a_2|}{\sqrt{v^2+a_{\perp}^2}},\quad  k_3=0.\label{nodalring}
\ee
For a general discussion of non-centrosymmetric nodal superconductors, see Refs.~\cite{Armitage2018,Schnyder,Chiu}.
Interaction-induced instabilities in nodal-line WSMs have also recently been studied, e.g., in Ref.~\cite{Volkov2018}.
  
  \begin{figure}
\begin{centering}
\includegraphics[width=\columnwidth]{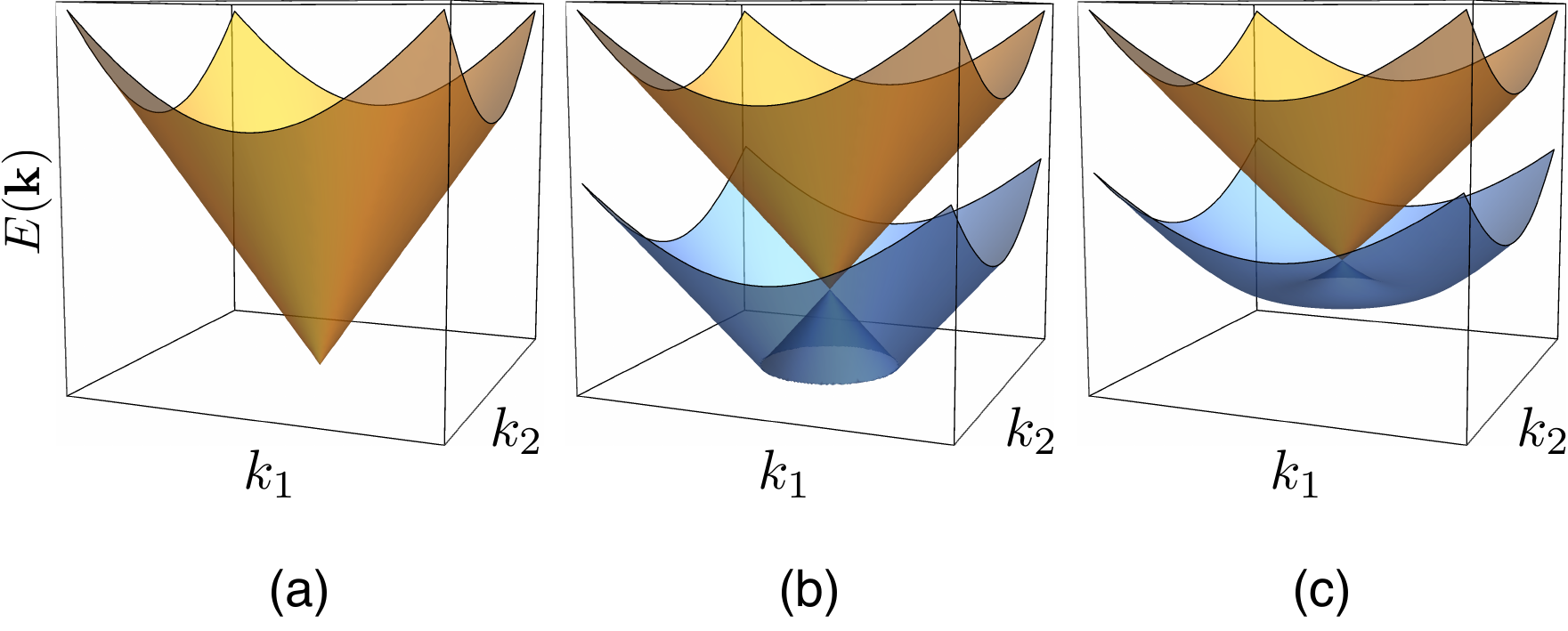}
\par\end{centering}
\caption{\label{fig7} Schematic representation of the dispersion relations  of  the two bands for Bogoliubov quasiparticles. Here we set the mean-field parameters $a_\parallel=a_\perp=0$  and plot the dispersion  for $k_3=0$. (a) For  $a_2=\Delta_0=0$, the  Weyl nodes conjugated by time reversal symmetry are represented as two degenerate Bogoliubov-Weyl nodes. (b) For $\Delta_0 =0$ but $a_2\neq0$, the spectrum is gapless along a nodal line located   in the $k_3=0$ plane. (c) For $a_2\neq0$ and $\Delta_0\neq0$, the spectrum is fully gapped. }
\end{figure}

The spectrum in Eq.~(\ref{spectrumnodal}) shows that the parameters $a_{\parallel}$ and $a_{\perp}$ mainly just renormalize Fermi velocities, without introducing essential new physics. In order to get tractable analytical expressions, we thus consider the case $a_{\parallel}=a_{\perp}=0$ in what follows.
In particular, we test whether it is energetically favorable to convert Weyl nodes into the nodal ring in Eq.~\eqref{nodalring} where the attractive interactions are most pronounced. To that end,  self-consistency equations for the order parameters are derived as shown in App.~\ref{app4}. We arrive  
at the coupled equations
\bea\label{eqa2}
a_2&=&\frac{\alpha_{\rm eff} a_2 }{4\pi} \int_0^\pi d\theta\sin\theta\left[\gamma(\theta)-1 \right]\\ \nonumber
&\times& \left(1+\sin^2\theta\right)\ln\left(\frac{4v^2b^2}{\Delta_0^2+a_2^2\cos^2\theta}\right),
\eea
and
\be\label{generalgapeq}
\Delta_0=\frac{\alpha_{\rm eff}\Delta_0}{4\pi} \int_0^\pi d\theta\sin\theta \left[\gamma(\theta)-1\right] \ln\left(\frac{4v^2b^2}{\Delta_0^2+a_2^2\cos^2\theta}\right).
\ee
Note that Eq.~\eqref{eqa2} differs from Eq.~\eqref{generalgapeq} by the factor $(1+\sin^2\theta)$ in the integrand. This factor enhances the contribution from $\theta\approx \pi/2$ where $\gamma(\theta)$ has its maximum. This observation suggests the existence of a parameter window  where Eq. (\ref{eqa2}) has a solution with $a_2\neq0$ while $\Delta_0=0$ is the only solution to Eq.~(\ref{generalgapeq}). 
In App.~\ref{app4}, we confirm that an intermediate parameter regime exists, $\bar\gamma'<\bar\gamma<1$, where such a solution is stable, at least in the absence of disorder.   Using TaAs parameters, we find $\bar\gamma'\simeq 0.91$. The respective value for the order parameter $a_2$ is given by Eq.~\eqref{a2gap}.  

Our mean-field approach suggest that superconductivity will be absent for $\bar\gamma< \bar\gamma'$, where the WSM phase presumably remains stable. We study the quasi-particle lifetime in this regime in Sec.~\ref{sec4c} below.
In the intermediate regime $\bar\gamma'<\bar\gamma<1$, however, the  system becomes a gapless triplet superconductor with inter-node pairing, where the Weyl nodes split and form a nodal ring located in the $k_3=0$ plane.
Finally, for $\bar\gamma>1$, the system enters a fully gapped superconducting phase with $s$-wave singlet pairing, see Sec.~\ref{sec4b1}.  The general picture is illustrated in Fig. \ref{fig7}.   We emphasize that all these phase transitions can
already happen for small absolute values of the   fine structure constant $\alpha=g_e^2/(4\pi   v)$, within the perturbatively accessible regime.

\subsubsection{Other competing phases}
\label{sec4b3}

So far we have discussed superconducting pairing with zero Cooper pair momentum in time-reversal-symmetric WSMs, where a pair of nodes at opposite momenta is conjugated by time reversal. 
By contrast, in inversion-symmetric WSMs, the opposite chirality of nodes entails that states with momentum $\mb k$ and $-\mb k$ do not necessarily have opposite spin. In such cases, the type of superconducting order is less clear because   pairing between parity-reversed nodes leads to a gapless superconductor \cite{Meng,Cho,Li}.   The authors of Refs.~\cite{Cho,Wei} have argued that a fully gapped Fulde-Ferrell-Larkin-Ovchinnikov (FFLO) state with intra-node pairing  has lower energy than the gapless state.  On the other hand, in Ref.~\cite{Bednik}   an odd-parity BCS state with lower energy than the FFLO state was found. Using our model, pairing between nodes of opposite chirality can also be studied and could allow for a nodal FFLO-type superconducting phase. 
However, paired states are then not related by any symmetry, and we find it unlikely that a lower energy than for the BCS state in Sec.~\ref{sec4b1} can be achieved for $\bar\gamma>1$. 
Moreover, our attractive phonon-mediated interaction favors pairing between time-reversal-conjugated nodes. In Eq.~\eqref{piezoqdep}, phonons couple to the total electronic density, and projecting $H_{\rm pz}$ onto the Weyl nodes at 
low energies, we find the same coupling to all nodes. Nonetheless, the process of integrating out high-energy modes could lift this degeneracy, and one pair of Weyl nodes may ultimately have a stronger coupling. 
The effective e-e interaction  used as input in Eq.~\eqref{heff1} will then favor a pairing of the time-reversal-conjugated nodes with the strongest coupling, 
as opposed to some other combination of nodes. 
 
 Let us also comment on the possibility of charge density wave (CDW) phases, see Ref.~\cite{Wang}. For the model with short-range attractive interactions in Ref.~\cite{Wang}, a CDW instability can only occur at strong coupling. It is straightforward to adapt their calculation to our model with long-range attraction. The mean-field Hamiltonian for the CDW state is essentially as for our singlet pairing state in Sec.~\ref{sec4b1}. The difference is that the four-component spinor is defined as $(\psi_{1\uparrow},\psi_{1\downarrow},\psi_{2\uparrow},\psi_{2\downarrow})^t$, where 1 and 2 now refer to two  nodes with opposite chirality and the order parameter is $\langle \psi^\dagger_1(\mb k)\psi^{\phantom\dagger}_2(\mb k)\rangle$. As this CDW order parameter breaks chiral symmetry, it leads to an axion insulator where the axion field is identified with the phase of the charge density wave. However, in our setting, this type of order  depends on the interaction between nodes which are not related by any symmetry. By the above argument, this state should have higher energy than the BCS state.
 
 In addition, there may be other phases at intermediate coupling strength, $\bar\gamma\lesssim 1$. One particularly intriguing possibility concerns phases that break time-reversal symmetry spontaneously, e.g., a $p+ip$ superconductor.  We leave the exploration of such phases to future work.

\section{Quasi-particle lifetime}\label{sec4c}
We next address the temperature and momentum dependence of the on-shell quasi-particle decay rate, $\Gamma(\mb p,T)$, caused by the piezoelectric e-ph coupling.
We assume that $\bar\gamma$ is so small that interaction-induced instabilities are absent. We show below that in this WSM phase, the
e-ph interaction is responsible for a rather large quasi-particle decay rate, scaling as $\Gamma\sim T/\ln(b/|\mb p|)$ at low-to-intermediate temperatures with $T\gg c_{ph}|\mb p|$. 
To ease notation, we again employ  Eqs.~\eqref{isotropic} and \eqref{isophon}.

\subsection{General expression for the decay rate}\label{sec4c1}

Diagrammatically, the lowest-order electronic self-energy is represented by Figs.~\ref{fig5}(a) and \ref{fig5}(d). Since the rainbow diagram in Fig.~\ref{fig5}(a) is a real-valued Hartree-Fock diagram, it does not contribute to the decay rate. The e-e interaction only produces a finite decay rate at higher orders and  beyond the Hartree-Fock approximation.  
In order to compute $\Gamma$, we therefore study the self-energy due to e-ph interactions, $\Sigma_{ep}$, see Fig.~\ref{fig5}(d). The rate follows from the 
imaginary part of $\Sigma_{ep}(E,\mb p)$, which in turn is obtained by analytic continuation $i\omega\to E+i0^+$, see, e.g., Ref.~\cite{Giraud2011}. 

To be specific, we study the lifetime of a Weyl quasi-particle in the state $|\mb p,\mu=+\rangle$ with momentum $\mb p$, taken from the positive-energy ($\mu=+$) band. 
We consider the on-shell case, $E=v|\mb p|$. The quasi-particle decay rate is then given by
\be \label{lifetimedef}
\Gamma ({\bf p}, T)= - 2 \  {\rm Im}\  \langle \mb p,+|\Sigma_{ep}(\mb p)|\mb p,+\rangle.
\ee
Let us now make use of the results of Sec.~\ref{sec3a4} and App.~\ref{app2}. 
We first observe that the decay rate must vanish right at the Weyl point, $\Gamma(\mb p=0,T)=0$, since then momentum and energy conservation cannot be satisfied for any phonon momentum ${\mb q}\ne 0$.
For $|\mb p|\ne 0$, it is convenient to rescale $\mb q=\xi |\mb p|  \hat {\mb q}$ with the dimensionless parameter $\xi$.
Denoting the integration angles by $\theta_{\mb q}$ and $\phi_{\mb q}$, and using
$\langle \mb p,+|\boldsymbol\sigma\cdot \mb q|\mb p,+\rangle=|\mb q| \hat {\mb q}\cdot \hat{\mb p},$ we find
\begin{widetext}
\bea
\Gamma({\mb p},T) &=&\frac{ g_e^2g_{ph}^2 c_{ph}^2 |\mb p|}{8\pi^2\rho_0} \int_0^\infty d\xi \,\xi^2\int_0^\pi d\theta_{\mb q}\sin\theta_{\mb q}\int_{-\pi}^{\pi}d\phi_{\mb q}\, \gamma(\hat{\mb q})  \label{decayrate}\sum_{s=\pm}  \Biggl \{ F_1^{(s)}(|\mb p|,\xi,\hat {\mb q}\cdot \hat{\mb p}) \times \\
&\times& \delta\left((v+s c_{ph} \xi)^2 -v^2\left(1+\xi^2+2\xi\hat{\mb q}\cdot\hat{\mb p}\right)\right) 
\nonumber + \ F^{(s)}_2(|\mb p|,\xi,\hat {\mb q}\cdot \hat{\mb p}) \ \delta\left(v^2\left (1-s\sqrt{1+\xi^2+2\xi\hat {\mb q}\cdot \hat{\mb p}}\right)^2-c_{ph}^2\xi^2\right) \Biggr\},
\eea
\bea  \label{Faux}
F_1^{(s=\pm)}&=& g^{(s)}_1(\xi) \frac{n_B\left(s c_{ph} |\mb p|\xi\right)}{s c_{ph} \xi}
\left[2v+s c_{ph} \xi+ v\xi\hat {\mb q}\cdot \hat{\mb p}\right],\\ \nonumber
F_2^{(s=\pm)}&=&-g^{(s)}_2(\xi,\hat {\mb q}\cdot \hat{\mb p}) \frac{n_F\left(sv|\mb p|\sqrt{
1+\xi^2+2\xi\hat {\mb q}\cdot \hat{\mb p}}\right)}{sv\sqrt{1+\xi^2+2\xi\hat {\mb q}\cdot \hat{\mb p}}}\left[v\sqrt{1+\xi^2+2\xi\hat {\mb q}\cdot \hat{\mb p}}+
sv\left(1+\xi \hat {\mb q}\cdot \hat{\mb p}\right)\right],
\eea
\end{widetext}
with $\gamma(\hat {\bm q})$ in Eq.~\eqref{phononpotential}, $n_F(\omega)=1/(e^{\beta\omega}+1)$, $n_B(\omega)=1/(e^{\beta\omega}-1)$,   and 
\bea
g_1^{(-)}&=&{\rm sgn}\left(1-\frac{c_{ph}\xi}{v}\right), \quad g_1^{(+)}=g_2^{(-)}=1, \\ \nonumber
\quad g_2^{(+)}&=&\textrm{sgn}\left(1-\sqrt{1+\xi^2+2\xi\hat{\mb q}\cdot\hat{\mb p}}\right).
\eea

\subsection{Zero-temperature limit}\label{sec4c2}

Let us first address the $T=0$ case, where only $F_1^{(-)}$ in Eq.~\eqref{Faux} yields a finite contribution to the decay rate,
\bea
&& \Gamma({\mb p}, T=0)=\frac{ g_e^2g_{ph}^2 c_{ph}|{\mb p}| }{4\pi^2\rho_0v}  \int_0^\pi d\theta_{\mb q}\sin\theta_{\mb q}\int_{-\pi}^{\pi}d\phi_{\mb q}
 \nonumber\\ \label{zerotemp} && \qquad \times \ 
\gamma(\hat{\mb q})\left[1-(\hat {\mb q}\cdot \hat {\mb p})^2 \right]\Theta(-\hat{\mb q}\cdot \hat {\mb p}),
\eea
where $\Theta(x)$ is the Heaviside step function.  Since the integral in Eq.~\eqref{zerotemp} is finite, we conclude that the $T=0$ rate scales as
$\Gamma \sim |\mb p|$ when approaching the Weyl point.

\subsection{Finite temperatures} \label{sec4c3}

Next we consider low but finite temperatures in the regime 
\be\label{Tregime}
 c_{ph}|\mb p|\ll T \ll \textrm{min}(v|\mb p|,c_{ph} b).
\ee
The dominant contributions to the decay rate \eqref{decayrate} then stem from the
$F_1^{(\pm)}$ terms in Eq.~(\ref{decayrate}), where the Bose factors can be approximated 
by $n_B\simeq \pm T/(c_{ph}|\mb p|\xi)$, respectively. We then obtain 
\bea\label{singular}
\Gamma({\mb p},T)&=& \frac{ g_e^2g_{ph}^2 T}{4\pi^2\rho_0 v} \int_0^\pi d\theta_{\mb q}\sin\theta_{\mb q}\int_{-\pi}^{\pi}d\phi_{\mb q}\\ \nonumber &\times& 
\frac{\gamma(\hat{\mb q})}{|\hat{\mb q}\cdot \hat{\mb p}|} \left[1-(\hat {\mb q}\cdot \hat{\mb p})^2\right]\Theta(-\hat{\mb q}\cdot \hat {\mb p}).
\eea 
However, the integral (\ref{singular}) diverges logarithmically at the boundary of the hemisphere 
$\hat {\mb q}\cdot \hat{\mb p}<0$, corresponding to small-angle scattering processes with $\xi\to 0$. 
This infrared divergence is related to the long-range character of the  piezoelectric interaction. 
Note that so far we have 
always assumed $T=0$, with the Fermi energy located right at the Weyl point. In that case, the unscreened Coulomb potential can be used.
For the finite-temperature quasi-particle decay rate, we need to be more careful since also finite-energy states within an
energy window of width $\approx T$ around the Weyl point are involved.  For such states, the long-range Coulomb interaction is 
modified by dynamic screening \cite{Throckmorton,Kozii}. By taking into account screening,  we now show that 
the above divergence is indeed removed.

Dynamic screening of the Coulomb interaction can be included by replacing the  permittivity according to \cite{MahanBook} 
\be
\varepsilon \mapsto \varepsilon(q) = \left( 1-\frac{g_e^2}{\mb q^2} \Pi(q) \right) \varepsilon,
\ee
where $\Pi(q)$ is the polarization function.  Within the standard random-phase approximation, we take $\Pi(q)$ to be the 
noninteracting polarization bubble, cf.~Fig.~\ref{fig5}(c), where  the $T=0$ limit of the 
polarization function yields a good description for the temperature regime \eqref{Tregime}.
A temperature dependence  of the decay rate is then generated only by e-ph interactions (we note that disorder effects could modify our expressions).
To obtain the dominant terms contributing to $\Gamma(\mb p,T)$ in this regime, the logarithmic on-shell term calculated in Refs.~\cite{Throckmorton,Abrikosov2} suffices,
\be
\Pi(q)\simeq -\frac{N |\mb q|^2}{6\pi^2 v}\ln\left (2b/|\mb q|\right),
\ee
where $b$ serves as large-momentum cutoff again. 
Note that two factors of $\varepsilon^{-1}$ appear in Eq.~\eqref{singular}, associated with either $g_e^2$ or $g_{ph}^2$. One can identify these 
two factors with the two wiggly lines  in the self-energy diagram in Fig.~\ref{fig5}(d). 
Dressing both lines with the polarization bubble, we arrive at a modified version of Eq.~(\ref{singular}) which takes into account screening,
\bea\label{finite}
&&\Gamma({\mb p},T)=\frac{g_e^2g_{ph}^2 T}{4\pi^2\rho_0 v} \int_0^\pi d\theta_{\mb q}\sin\theta_{\mb q}\int_{-\pi}^{\pi}d\phi_{\mb q}\\
&&\quad \times  \nonumber \frac{\gamma(\hat{\mb q})}{|\hat {\mb q}\cdot \hat{\mb p}|}
\frac{1-( \hat {\mb q}\cdot \hat{\mb p})^2 }{\left[1+\frac{N g_e^2}{6\pi^2 v}\ln\left(\frac{1}{|\hat{\mb q}\cdot \hat {\mb p}|}\frac{b}{|\mb p|}\right)
\right]^2}  \Theta(-\hat{\mb q}\cdot \hat {\mb p}).
\eea
Using $|\mb p|\ll b$, the regime \eqref{Tregime} is therefore characterized by a quasi-particle decay rate which scales as
\be\label{decayratefinal}
\Gamma(\mb p,T)\sim \frac {T}{\ln(b/|\mb p|)} .
\ee
We observe that $\Gamma({\mb p},T)$ vanishes for $|\mb p|\to 0$, as expected from  kinematic constraints. However, 
the slow logarithmic scaling with $|\mb p|$, together with the linear-$T$ dependence, suggests that the quasi-particle 
lifetime of Weyl fermions is significantly reduced by the piezoelectric e-ph interaction, even when one stays in the very close vicinity of a Weyl point.

\section{Concluding remarks}\label{sec5}

In this work we have studied the long-range attractive interactions mediated by the piezoelectric electron-phonon coupling in undoped non-centrosymmetric Weyl semimetals.  These interactions exhibit a significant angular dependence and compete with the  repulsive Coulomb interactions.  This competition is mainly governed by the dimensionless piezoelectric coupling strength $\bar\gamma$ in Eq.~\eqref{bargamma}.  Within a static approximation for the effective e-e interaction, we find that for $\bar\gamma>1$ the attractive interactions outweigh the repulsive Coulomb part. We then predict a conventional BCS superconductor phase with spin-singlet $s$-wave pairing, even though the normal density of states vanishes at the Fermi level. 
We have performed a mean-field analysis to study this state in some detail. 

According to our rough estimate  $\bar\gamma\approx 0.20$ for TaAs, see Sec.~\ref{sec4a},  the above BCS  scenario is probably hard to encounter in TaAs.  However, for $\bar\gamma<1$, other, and even more interesting,  interacting phases may be stabilized. For example, our analysis in Sec.~\ref{sec4b} suggests that a nodal-ring gapless spin-triplet superconductor will be realized for intermediate values of $\bar \gamma$.  Our RG analysis also shows that the critical values for $\bar \gamma$ where superconducting instabilities are found can be pushed upwards by retardation effects.

For small $\bar\gamma$, we expect that the Weyl semimetal phase remains stable. Nonetheless, the  piezoelectric coupling should leave a clear experimental trace in the quasi-particle decay rate at finite temperature. In particular,  we find that this rate scales as $\Gamma\sim T/\ln (b/|\mb p|)$ at low-to-intermediate $T$.   Albeit $\Gamma=0$ right at a Weyl point ($\mb p=0$), the weak logarithmic scaling with $|\mb p|$ suggests that the quasi-particle lifetime will be rather short even for very small (but finite) $|\mb p|$.
In any case, we hope that future theoretical and experimental research will continue to study the interesting consequences of piezoelectric  couplings in Weyl semimetals.

\begin{acknowledgements}
We acknowledge funding by the Deutsche Forschungsgemeinschaft (DFG), Grant No.~EG 96/12-1. R.G.P. thanks the Humboldt foundation for a Bessel award, enabling his extended stay in D\"usseldorf. Research at IIP-UFRN is supported by the Brazilian ministries MEC and MCTIC.
\end{acknowledgements}

\appendix

\section{Electronic self-energy from Coulomb interactions}\label{app1}

Here we provide details concerning the electronic self-energy correction from Coulomb interactions, see Sec.~\ref{sec3a1}.
Using Eq.~\eqref{GF1} and performing the internal frequency integration in Eq.~\eqref{selfenergy1}, we obtain
\bea\label{B1}
\Sigma_{ee}(i\omega,\mb p)&=&\frac{ g_e^2}{2}\int\frac{d^3\mb q}{(2\pi)^3} \frac{1}{\mb q^2} \Biggl( -\sigma_0 + \\
\nonumber &+& \frac{v_\perp(\mb p_\perp+\mb q_\perp)\cdot \boldsymbol \sigma_\perp+v_3(p_3+q_3)\sigma_3}{E(\mb p+\mb q)}
\Biggr),
\eea
which is independent of the frequency $\omega$. We now expand Eq.~\eqref{B1} for $|\mb p|\ll |\mb q|$. For $\mb p=0$, one finds a
non-universal constant that can be absorbed by renormalization of the chemical potential. The renormalized value of the chemical potential is then assumed to be aligned with the Weyl node.
Universal RG contributions appear at the first order in $\mb p$. For instance, from terms linear in $p_3$, we get a contribution of the form
\be
\Sigma_{ee}(p_3)= \frac{g_e^2}{2} v_3 p_3\sigma_3\int\frac{d^3\mb q}{(2\pi)^3}\frac{v_\perp^2\mb q_\perp^2}{\mb q^2 E^3(\mb q)},
\ee
where momentum-shell integration yields the self-energy correction 
\bea\label{B2}
\delta \Sigma_{ee}(p_3)&=&\frac{g_e^2}{8\pi^2} \eta_3  p_3\sigma_3 d\ell,\\  \nonumber
\eta_3&=&\frac{v_3}{v_\perp}\int_0^\pi d\theta \frac{\sin^3\theta}{[\sin^2\theta+(v_3/v_\perp)^2\cos^2\theta]^{3/2}}.
\eea
Self-energy terms $\sim\mb p_\perp$ follow in a similar manner with $\eta_3$  replaced by $\eta_\perp=4v_\perp/(3v_3)$.  
For the isotropic case \eqref{isotropic}, we then find $\eta_\perp=\eta_3=4/3$. The complete linear-in-$\mb p$ self-energy correction after  momentum-shell integration is given by Eq.~\eqref{selfenergy2}. 

\section{Electron self-energy from e-ph interactions}\label{app2}

Next we turn to the  self-energy  $\Sigma_{ep}(p)$ due to piezoelectric interactions, see Sec.~\ref{sec3a4}.  The leading term arises from the diagram in Fig.~\ref{fig5}(d),
\bea\label{C1}
\Sigma_{ep}(p)&=&-g_e^2g_{ph}^2 \sum_{ijk}\sum_{lmn}  e_{ijk}  e_{lmn}\\ \nonumber
&\times&
\int\frac{d^4q}{(2\pi)^4}\frac{q_iq_jq_lq_m}{|\mb q|^4}  D_{kn}(q)\mathbb G(p+q),
\eea
with the phonon propagator in Eq.~\eqref{Djjphonon} and the electronic GF in Eq.~\eqref{GF1}.
Performing the internal frequency integration, we obtain 
\begin{widetext}
\bea
\Sigma_{ep}(i\omega,\mb p)&=&-\frac{g_e^2 g_{ph}^2}{4\rho_0}\sum_{J}\sum_{ijk}\sum_{lmn}  e_{ijk}  e_{lmn}\int\frac{d^3\mb q}{(2\pi)^3}\frac{q_iq_jq_lq_m   \epsilon^J_{k}(\mb q)  \epsilon^J_{n}(-\mb q)}{|\mb q|^4\Omega_J(\mb q)E(\mb p+\mb q)}\nonumber\\
&\times& \label{B4} \sum_\pm \frac{[i\omega\pm\Omega_J(\mb q)]\sigma_0+v_\perp(\mb p_\perp+\mb q_\perp)\cdot \boldsymbol \sigma_\perp+v_3(p_3+q_3)\sigma_3}{\pm i\omega+\Omega_J(\mb q)+ E(\mb p+\mb q)}.
\eea
\end{widetext}
We now integrate out all phonon modes within the high-energy shell.  
For $4mm$ crystal symmetry and using Eqs.~\eqref{isotropic} and \eqref{isophon}, 
to linear order in $p$, we find the correction
\be \label{selfaux}
\delta\Sigma_{ep}(p)=-g_e^2 \left(\frac{i\zeta_0\omega}{v} \sigma_0 +\zeta_\perp \mb p_\perp\cdot\boldsymbol\sigma_\perp+\zeta_3p_3\sigma_3\right)d\ell ,
\ee
with, cf.~Eq.~\eqref{AB},
\bea\label{zeta0}
\zeta_0&=&\frac{C_0}{\rho_0} \left( \frac{g_{ph} e_{33}}{c_{ph}}\right)^2 \frac{c_{ph}}{v},
\\ \nonumber
C_0&=&\frac1{15\pi}\left(10 A^2+4AB+4A+2 B^2+3  \right) \simeq 1.40.
\eea
The terms $\sim \mb p$ in Eq.~\eqref{selfaux} involve dimensionless numbers $\zeta_\perp$ and $\zeta_3$ defined 
as in Eq.~\eqref{zeta0} but with $C_0\to C_{\perp,3}$, where 
\bea
C_\perp&=& \frac{1}{105\pi } \left(14 A^2+12AB +12 A+6 B^2+15 \right),\nonumber\\
C_3&=&   \frac{1}{105\pi}\left(42 A^2+4AB +4 A  +2 B^2-9 \right).
\eea
Using Eq.~\eqref{AB} with the parameters quoted in section \ref{sec4a}, we find $C_\perp\simeq 0.29$ and $C_3\simeq 0.83$.
The self-energy corrections are summarized in Eq.~\eqref{selfenergy3}.

\section{Vertex corrections} \label{app3}

Here we provide details about the vertex correction due to the diagram in Fig.~\ref{fig5}(e), see Sec.~\ref{sec3a5}.
We define the three-point function with fermions in the band basis \be
\Lambda_{\mu \mu'}(x,x',x'')=\langle T\psi^{\phantom\dagger}_\mu (x)\psi^\dagger_{\mu'}(x')\varphi(x'')\rangle.
\ee
The free boson propagator is $D_\varphi(x)=\langle \varphi(x)\varphi(0)\rangle$. In momentum space, $
D_\varphi(q)=-|\mb q|^{-2}$.

\begin{widetext}
The three-point function at tree level is obtained by perturbation theory to first order in $g_e$. Taking the Fourier transform $\Lambda_{\sigma\sigma' l}(p,p',q)=\int d^4 xd^4x' d^4x''\, e^{-ipx-ip'x'-iqx''}\Lambda_{\sigma\sigma' l}(x,x',x'')$, we find  \bea
\Lambda^{(1)}_{\mu \mu'}(p,p',q)&=&ig_e  [\mc U^\dagger(\mb p)\mc U(\mb p+\mb q)]_{\mu\mu'}G_\mu (p)  G_{\mu'}(p+q)  D_\varphi(q)(2\pi)^4\delta(p'+p+q).
\eea

The vertex correction at the one-loop level appears at order $g_e^3g_{ph}^2$:\bea
\Lambda^{(3,2)}_{\mu \mu'}(p,p',q)&=&-ig_e^3g_{ph}^2D_\varphi(q)(2\pi)^4\delta(p'+p+q)\sum_{amb}\sum_{rns}    e_{amb}  e_{rns}  \nonumber\\
&&\times\sum_{\mu''\mu'''}\int\frac{d^4 q'}{(2\pi)^4}  \frac{q_a'q_b'q_r'q_s'}{|\mb q'|^4}D_{mn}(q')G_\mu(p)  G_{\mu''}(p-q')  G_{\mu'''}(p+q-q')  G_{\mu'}(p+q)\nonumber\\
&&\times  [\mc U^\dagger(\mb p)\mc U(\mb p-\mb q')]_{\mu\mu''}  [\mc U^\dagger(\mb p-\mb q')\mc U(\mb p+\mb q-\mb q')]_{\mu''\mu'''} [\mc U^\dagger(\mb p+\mb q-\mb q')\mc U(\mb p+\mb q)]_{\mu'''\mu'}, 
\eea
with the phonon propagator  in Eq.~\eqref{phonD} with $q'=(i\nu',\mb q')$.

At low energies, all external four-momenta can be assumed small against $q'$. 
The integral over the internal frequency $\nu'$ then defines the quantities
\be
I_{\mu''}=\int_{-\infty}^{+\infty} \frac{d\nu'}{2\pi}\, \left(\frac{1}{i\nu'-\Omega_J(\mb q')}-\frac{1}{i\nu'+\Omega_J(\mb q')}\right)\frac{1}{i\nu -i\nu'-\mu'' E(\mb p-\mb q')}
\frac{1}{i\nu+i\omega -i\nu'-\mu'' E(\mb p+\mb q-\mb q')}.\label{integral}
\ee
Performing the frequency integral, we obtain
\be
I_{\pm}=-\frac{1}{[i\nu \mp \Omega_J(\mb q')\mp E(\mb p-\mb q')][i\nu+i\omega \mp \Omega_J(\mb q')\mp E(\mb p+\mb q-\mb q')]}.
\ee
 We now effectively set the external four-momenta to zero, $p,q\to 0$, which results in  
$I_{\pm}=-1/[\Omega_J(\mb q')+E(\mb q')]^2$.
In this limit, ${\mc U(\mb p-\mb q')\mc U^\dagger(\mb p+\mb q-\mb q')}\to \sigma_0$,
and we arrive at
\bea  \label{A8} 
\Lambda^{(3)}_{\mu\mu' l}(p,p',q)&=&\frac{ig_e^3 g_{ph}^2}{2\rho_0}   [\mc U^\dagger(\mb p)\mc U(\mb p+\mb q)]_{\mu\mu'}G_\mu(p)     G_{\mu'}(p+q) D_\varphi(q) (2\pi)^4\delta(p'+p+q)\\ \nonumber
&\times&\sum_J\sum_{amb}\sum_{rns}  e_{amb}  e_{rns}  \int\frac{d^3 \mb q'}{(2\pi)^3}  \frac{q_a'q_b'q_r'q_s'}{|\mb q'|^4}  \frac{  \epsilon^J_m(\mb q')  \epsilon^J_n(-\mb q')}{\Omega_J(\mb q')[\Omega_J(\mb q')+E(\mb q')]^2}.
\eea
At this point, we compute the one-loop contribution to the RG equations by integrating over phonon modes with momenta within the high-energy momentum shell. The correction corresponds to a charge renormalization, and hence to a renormalization of the coupling $g_e$ in Eq.~\eqref{gedef}. We find 
\be
\delta g_e=\frac{ g_e^3 g_{ph}^2 }{2\rho_0}  \sum_J  \int\frac{d^3  \mb q}{(2\pi)^3}  \left|\sum_{jkl}   e_{jkl}  \frac{q_jq_k}{|\mb q|^2}\epsilon^J_l(\mb q)\right|^2 \frac1{\Omega_J(\mb q)[\Omega_J(\mb q)+E(\mb q)]^2}.
\ee
\end{widetext}
 
For the case of TaAs, with the simplifications in Eqs.~\eqref{isotropic} and \eqref{isophon},
one can then employ similar steps as in Apps.~\ref{app1} and \ref{app2}. We thereby arrive at Eq.~\eqref{deltacharge}.

\section{On triplet pairing}\label{app4}

In this appendix, we provide details concerning the solution of 
the self-consistency equations in Sec.~\ref{sec4b2} for the triplet pairing case, keeping only $\Delta_0$ and $a_2$ as free parameters. The self-consistency equations are given by
\bea\label{E1}
\Delta_0&=&-\frac1{2V}\sum_{\mb k}V_{\text{tot}}(\mb k)\textrm{Tr}[{\bm\Xi}(\mb k)],\\ \nonumber
a_2&=&-\frac1{2V}\sum_{\mb k}V_{\text{tot}}(\mb k)\textrm{Im}\textrm{Tr}[{\bm\Xi}(\mb k)\sigma_3]. 
\eea 
Now suppose that ${\cal H}_{\rm BdG}(\mb k)$ is diagonalized by the unitary transformation $\Psi(\mb k)=U(\mb k)\tilde{\Psi}(\mb k)$,
with eigenvalues ordered as $( E_1(\mb k),- E_1(\mb k),- E_2(\mb k),E_2(\mb k))$. For an arbitrary $2\times 2$ matrix ${\bm W}$, we can use the auxiliary relation
\be
\textrm{Tr}[{\bm\Xi}(\mb k) W]=-\sum_{\lambda=2,3}[U^\dagger(\mb k)\tau^- WU(\mb k)]_{\lambda\lambda},
\ee
where $\tau^\pm=\tau_x\pm i\tau_y$ and Pauli matrices $\tau_{x,y,z}$ act in Nambu space.
The self-consistency equations (\ref{E1}) then reduce to
\bea
\Delta_0&=&\frac1{2}\sum_{\lambda=2,3}\int\frac{d^3\mb k}{(2\pi)^2}V_{\text{tot}}(\mb k)[U^\dagger(\mb k)\tau^-\sigma_0U(\mb k)]_{\lambda\lambda},\\ \nonumber
a_2&=&\frac1{2}\sum_{\lambda=2,3}\int\frac{d^3\mb k}{(2\pi)^2}V_{\text{tot}}(\mb k)\textrm{Im}[U^\dagger(\mb k)\tau^-\sigma_3U(\mb k)]_{\lambda\lambda}, 
\eea
where the integration domain is the ball $0<|\mb k|<b$. 
Defining the function 
\be
R(x,\theta)=\sum_\pm  \frac{a_2\pm x\sin\theta}{\sqrt{a_2^2\mp 2 a_2 x \sin \theta +\Delta_0 ^2+x^2}},
\ee
the above equation for $a_2$ takes the form
\be
a_2=\frac{\alpha}{4\pi}\int_0^{vb}dx\int_0^\pi d\theta\sin\theta\  R(x,\theta)\ [\gamma(\theta)-1],
\ee
where $\alpha=g_e^2/(4\pi v)$.
Performing the integral over $x$ and assuming  $a_2,\Delta_0  \ll vb$, we obtain
Eq.~\eqref{eqa2}.
In a similar fashion, we obtain the equation for $\Delta_0$ in Eq.~\eqref{generalgapeq}.

Setting  $\Delta_0=0$ and substituting the form of $\gamma(\theta)$ in Eq.~(\ref{ff2}), 
we get from Eq.~\eqref{eqa2}  
\bea
1&=&\frac{\alpha }{2\pi} \int_0^\pi d\theta\sin\theta[\gamma(\theta)-1]\left(1+\sin^2\theta\right)\ln\left|\frac{2vb}{a_2\cos\theta}\right|\nonumber\\
&=&-\frac{\alpha }{2\pi}\left[L_1(\bar\gamma)+L_2(\bar\gamma)\ln\left(\frac{a_2}{2vb}\right)\right].
\eea
We here define the linear functions $L_{1,2}(\bar\gamma)$ as
\begin{widetext}
\bea
L_1(\bar\gamma)&=&\frac{34}{9}-\frac{\bar\gamma}{w_\gamma}\left(\frac{256 A^2 }{75}+\frac{5408 A B }{11025}+\frac{5408 A }{11025}+\frac{2704 B^2 }{11025}+\frac{146 }{1225}\right),\nonumber\\
L_2(\bar \gamma)&=&-\frac{10}{3}+\frac{\bar\gamma}{w_\gamma}\left(\frac{18 }{35}+\frac{1}{5} 12 A^2 -\frac{88 A B }{105}-\frac{88 A }{105}-\frac{44 B^2 }{105}\right),
\eea
\end{widetext} 
with $w_\gamma\simeq 4.40$ in Eq.~\eqref{CPi}.
Solving for $a_2$, we obtain
\be \label{a2gap}
a_2=2vb \exp\left(-\frac{2\pi +\alpha L_1}{\alpha L_2}\right).
\ee
For $\alpha\lesssim 1$ and $\bar \gamma\lesssim 1$, the constraint $a_2<2vb$ simplifies to $L_2(\bar\gamma)>0$. In order to obtain a nontrivial solution for $a_2$, this in turn requires that  
\be \label{gcprime}
\bar \gamma>\bar\gamma'= \frac{175 w_\gamma}{126 A^2+44 A B+44 A+22 B^2+27}.
\ee
For TaAs parameters, Eq.~\eqref{gcprime} yields $\bar\gamma' \simeq 0.91$.

\end{document}